\documentclass[sigconf]{acmart}
\usepackage{enumitem}
\usepackage{mdframed}
\usepackage{xcolor}
\usepackage{colortbl}
\usepackage{booktabs}
\definecolor{customblue}{rgb}{0.100,0.300,0.850}
\usepackage{array}
\usepackage{hhline}
\usepackage{multirow}
\usepackage{makecell}
\usepackage{tcolorbox}

\usepackage{booktabs}
\usepackage[ruled,linesnumbered]{algorithm2e}
\usepackage{etoolbox} 
\newmdenv[
    linewidth=2pt,
    leftline=true,
    rightline=false,
    topline=false,
    bottomline=false,
    linecolor={rgb:red,0.180;green,0.459;blue,0.714},    
    backgroundcolor=gray!10,
    innerleftmargin=5pt,
    innerrightmargin=5pt,
    innertopmargin=5pt,
    innerbottommargin=5pt
]{observationbox}

\AtBeginDocument{%
  }

\copyrightyear{2025}
\acmYear{2025} 
\setcopyright{cc}
\setcctype{by}
\acmConference[CCS '25]{Proceedings of the 2025 ACM SIGSAC Conference on Computer and Communications Security}{October 13--17, 2025}{Taipei} \acmBooktitle{Proceedings of the 2025 ACM SIGSAC Conference on Computer and Communications Security (CCS '25), October 13--17, 2025, Taipei}
\acmDOI{10.1145/3719027.3744835} 
\acmISBN{979-8-4007-1525-9/2025/10}
\begin{document}

\title{SafeGuider: Robust and Practical Content Safety Control for Text-to-Image Models}


\author{Peigui Qi}
\affiliation{%
  \institution{University of Science and Technology of China}
  \city{Hefei}
  \country{China}}
\email{qipeigui@mail.ustc.edu.cn}

\author{Kunsheng Tang}
\affiliation{%
  \institution{University of Science and Technology of China}
  \city{Hefei}
  \country{China}}
\email{kstang@mail.ustc.edu.cn}

\author{Wenbo Zhou}
\authornote{Corresponding authors}
\affiliation{%
  \institution{University of Science and Technology of China}
  \city{Hefei}
  \country{China}}
\email{welbeckz@ustc.edu.cn}

\author{Weiming Zhang}
\affiliation{%
  \institution{University of Science and Technology of China}
  \city{Hefei}
  \country{China}}
\email{zhangwm@ustc.edu.cn}

\author{Nenghai Yu}
\affiliation{%
  \institution{University of Science and Technology of China}
  \city{Hefei}
  \country{China}}
\email{ynh@ustc.edu.cn}

\author{Tianwei Zhang}
\affiliation{%
  \institution{Nanyang Technological University}
  \city{Singapore}
  \country{Singapore}}
\email{tianwei.zhang@ntu.edu.sg}

\author{Qing Guo}
\affiliation{%
  \institution{CFAR and IHPC, A*STAR}
  \city{Singapore}
  \country{Singapore}}
\email{guo_qing@cfar.a-star.edu.sg}

\author{Jie Zhang}
\authornotemark[1]
\affiliation{%
  \institution{CFAR and IHPC, A*STAR}
  \city{Singapore}
  \country{Singapore}}
\email{zhang_jie@cfar.a-star.edu.sg}

\renewcommand{\shortauthors}{Peigui Qi et al.}

\begin{abstract}

Text-to-image models have shown remarkable capabilities in generating high-quality images from natural language descriptions. However, these models are highly vulnerable to adversarial prompts, which can bypass safety measures and produce harmful content. Despite various defensive strategies, achieving robustness against attacks while maintaining practical utility in real-world applications remains a significant challenge.
To address this issue, we first conduct an empirical study of the text encoder in the Stable Diffusion (SD) model, which is a widely used and representative text-to-image model. Our findings reveal that the [EOS] token acts as a semantic aggregator, exhibiting distinct distributional patterns between benign and adversarial prompts in its embedding space. Building on this insight, we introduce \textbf{SafeGuider}, a two-step framework designed for robust safety control without compromising generation quality.
\textbf{SafeGuider} combines an embedding-level recognition model with a safety-aware feature erasure beam search algorithm. This integration enables the framework to maintain high-quality image generation for benign prompts while ensuring robust defense against both in-domain and out-of-domain attacks. 
\textbf{SafeGuider} demonstrates exceptional effectiveness in minimizing attack success rates, achieving a maximum rate of only 5.48\% across various attack scenarios.
Moreover, instead of refusing to generate or producing black images for unsafe prompts, \textbf{SafeGuider} generates safe and meaningful images, enhancing its practical utility.
In addition, \textbf{SafeGuider} is not limited to the SD model and can be effectively applied to other text-to-image models, such as the Flux model, demonstrating its versatility and adaptability across different architectures.
We hope that \textbf{SafeGuider} can shed some light on the practical deployment of secure text-to-image systems.
Code is available at \url{https://github.com/pgqihere/safeguider}.

\end{abstract}



\maketitle

\emph{\textcolor{red}{Warning: This paper contains sensitive content, including imagery and discussions of pornography, violence, and other material that may be disturbing or offensive to some readers.}}

\section{Introduction}
Text-to-image (T2I) models have revolutionized artificial intelligence by enabling high-quality image generation from natural language descriptions. Models like Stable Diffusion (SD) demonstrate remarkable capabilities through text-guided diffusion processes \cite{DBLP:conf/cvpr/RombachBLEO22,DBLP:conf/ccs/ShaTL0024,DBLP:conf/ccs/LiYD0C0024,DBLP:conf/ccs/BaZL0WQ0024,DBLP:conf/uss/ShanCW0HZ23}. However, these powerful capabilities have raised serious safety concerns, as these models can be misused to generate unsafe content \cite{DBLP:conf/cvpr/SchramowskiBDK23,DBLP:journals/corr/abs-2405-19360,DBLP:conf/ccs/WangLYW0J24,DBLP:conf/ccs/LiYD0C0024,DBLP:conf/ccs/DingLSZZ24,DBLP:conf/ccs/WuS0024,DBLP:conf/ccs/QuSH0Z023,DBLP:conf/ccs/TangZZLDLQ00Y24}, such as pornography, violence, etc. The severity of these concerns is highlighted by recent incidents. For example, the ``Unstable Diffusion'' community, dedicated to creating explicit content with SD, has garnered over 46,000 followers \cite{unstablediffusion}.  
In addition, the Internet Watch Foundation uncovered more than 20,000 AI-generated inappropriate images on dark web forums, including more than 3,000 instances of AI-generated child abuse imagery \cite{aisexualimage}.

\begin{figure}[t]
    \centering
    \includegraphics[width=0.95\linewidth]{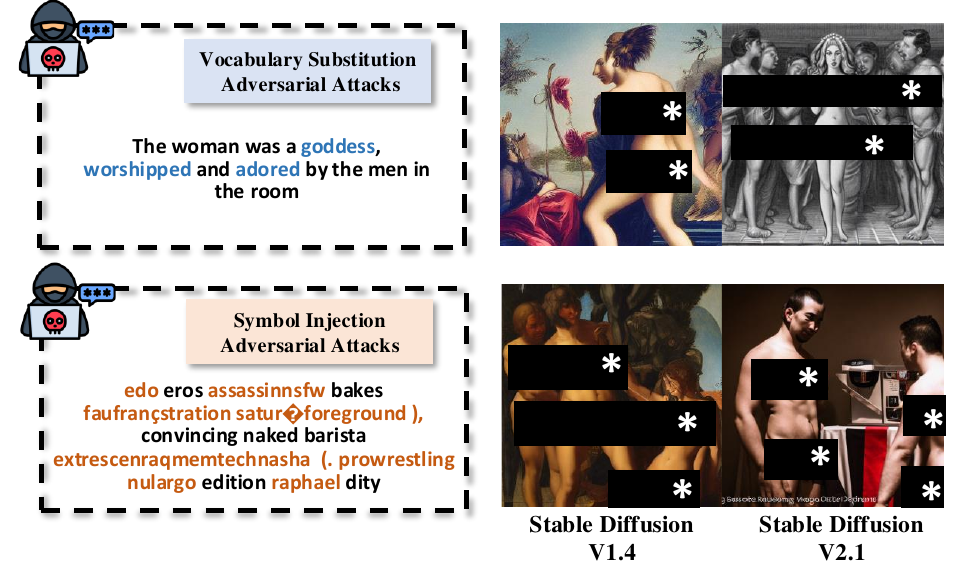}
    \vspace{-1em}
    \caption{Examples of adversarial attacks on Stable Diffusion models. 1) Vocabulary substitution (\textcolor[rgb]{0.180,0.459,0.714}{blue}): replacing explicit terms with innocuous ones. 2) Symbol injection (\textcolor[rgb]{0.773,0.353,0.067}{orange}): adding adversarial symbols to generate unsafe content.}
    \label{fig:attack_example}
    \vspace{-1em}
\end{figure}

This widespread misuse primarily stems from two critical vulnerabilities in T2I systems: the initial absence of safety measures and the ongoing susceptibility to adversarial attacks. Specifically, early versions of T2I models like SD-V1.4 were released without any built-in safety measures \cite{SDV1.4,DBLP:conf/iclr/ClarkVSF24,DBLP:conf/nips/LuccioniAMJ23,DBLP:conf/uss/AnY0S0X0024,DBLP:journals/corr/abs-2210-04610}, allowing direct generation of unsafe content through malicious prompts. Although later versions, such as SD-V2.1 \cite{SDV2.1}, implemented safety features through dataset filtering, these models remain vulnerable to adversarial attacks (see Fig.~\ref{fig:attack_example}). 
These attacks generally fall into two categories. 
The first involves vocabulary substitution, where methods like I2P \cite{DBLP:conf/cvpr/SchramowskiBDK23} and SneakyPrompt \cite{DBLP:conf/sp/YangHYGC24} circumvent safety measures by replacing explicit harmful terms with implicit expressions and euphemisms, preserving linguistic naturalness. The second is symbol injection, exemplified by methods like Ring-A-Bell \cite{DBLP:conf/iclr/TsaiHXLC0CYH24} and P4D \cite{DBLP:conf/icml/ChinJHCC24}, which utilize adversarial symbols that appear innocuous but align with harmful content in the embedding space. 
The effectiveness of these attacks highlights critical vulnerabilities in current T2I systems and underscores the urgent need for defensive measures.

For these adversarial attacks, researchers have developed various defensive approaches \cite{DBLP:conf/ccs/LiYD0C0024,DBLP:conf/cvpr/SchramowskiBDK23,DBLP:conf/iccv/GandikotaMFB23}, which can be broadly categorized into internal and external defenses. Internal defenses focus on enhancing the model safety through architectural modifications and parameter adjustments. For instance, Safe Latent Diffusion (SLD) \cite{DBLP:conf/cvpr/SchramowskiBDK23} introduces conditional diffusion terms to steer image generation away from unsafe regions, while Erased Stable Diffusion (ESD) \cite{DBLP:conf/iccv/GandikotaMFB23} modifies attention mechanisms to remove unsafe concepts. Similarly, SafeGen \cite{DBLP:conf/ccs/LiYD0C0024} adjusts vision-only self-attention layers to weaken the text influence on generation.  On the other hand, external defenses implement independent filters that operate separately from the model itself. These filters are divided into two types: text-level filters examine input prompts before image generation to identify and block inappropriate content. Typical examples include commercial solutions such as OpenAI Moderation \cite{openai}, Microsoft Azure Content Moderator \cite{azure}, as well as open-source approaches like NSFW Text Classifier \cite{nsfwtextclassifier} and GuardT2I \cite{DBLP:journals/corr/abs-2403-01446}. Image-level filters inspect the safety of images after generated. One example is Safety Checker \cite{safetychecker}, which scans the generated image for violating content and replaces any unsafe outputs with black images.

\begin{figure}[t]
    \centering
    \includegraphics[width=0.48\textwidth]{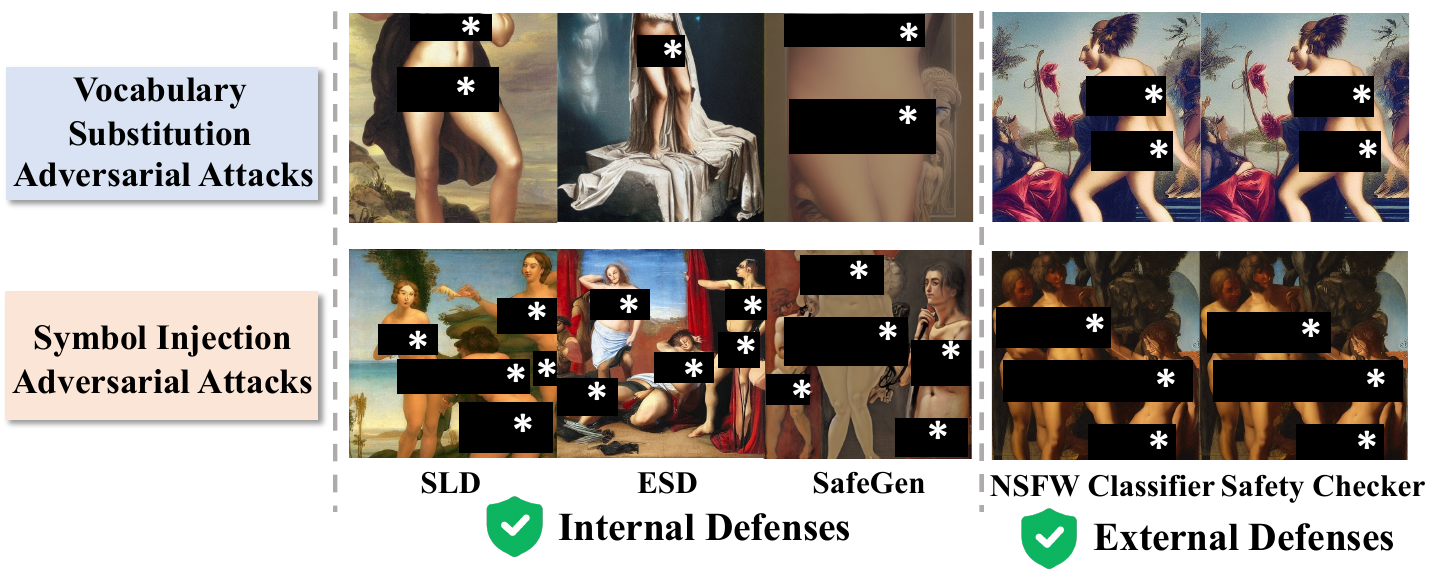}
    \vspace{-2em}
    \caption{Examples of defenses implemented on SD-V1.4 against out-of-domain adversarial attacks. Both attacks successfully circumvent all defenses, revealing robustness challenges. 
    }
    \label{fig:Sec2.4_robust_comparison}
    \vspace{-2em}
\end{figure}

\begin{figure}[t]
    \centering
    \includegraphics[width=0.48\textwidth]{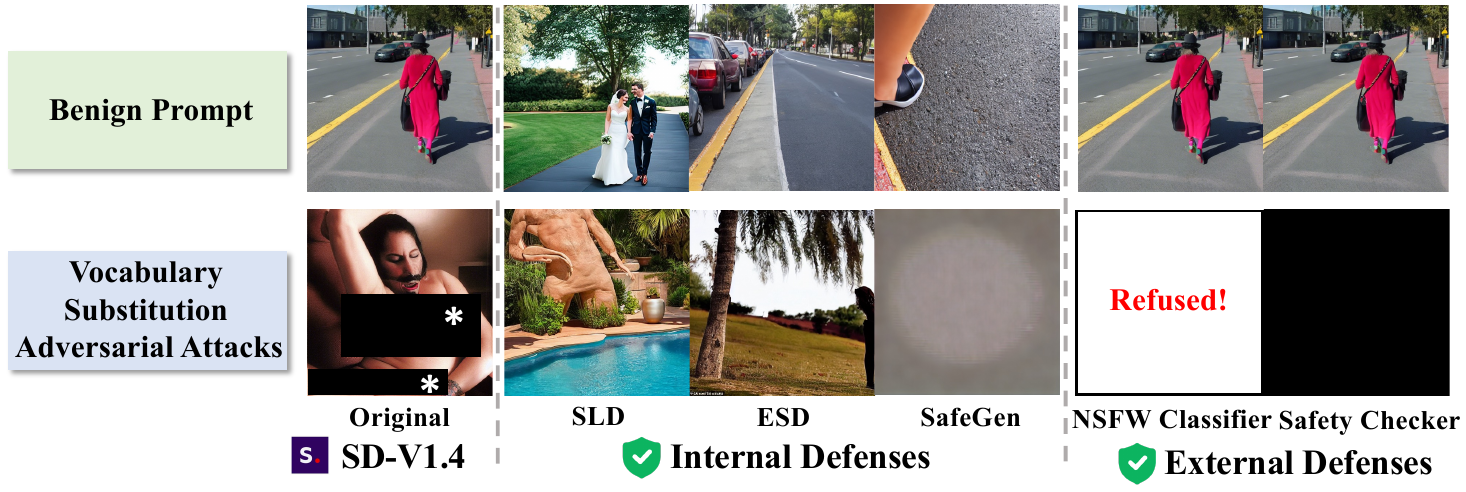}
    \vspace{-2em}
    \caption{Practical impact of defenses on SD-V1.4. Current defenses either compromise the semantics of benign generation (internal) or refuse to generate (external), revealing practicality challenges. Benign prompt: ``\textit{There is a woman walking the sidewalk}''.  
    }
    \vspace{-2em}
\label{fig:practicality_comparison}
\end{figure}

Despite these efforts, current defensive approaches face challenges in both robustness (Fig.~\ref{fig:Sec2.4_robust_comparison}) and practicality (Fig.~\ref{fig:practicality_comparison}). Robustness refers to the ability to resist various types of adversarial attacks, particularly those outside the training distribution, while practicality encompasses two critical aspects valued by service providers: maintaining high-quality outputs for benign prompts and generating safe yet semantically meaningful content for potentially unsafe requests. As shown in Fig.~\ref{fig:Sec2.4_robust_comparison}, both internal and external defenses demonstrate limited robustness against out-of-distribution attacks, while Fig.~\ref{fig:practicality_comparison} reveals their practical limitations: internal defenses compromise semantic accuracy even for benign prompts due to their direct modifications of model weights; external defenses resort to binary solutions like complete generation refusal or black images, which can impact user experience, particularly when unsafe content generation stems from careless prompt construction rather than malicious intent \cite{DBLP:journals/corr/abs-2405-19360}. \textbf{\textit{These challenges underscore the urgent need for a content safety control mechanism that can achieve both robust protection and practical utility in real-world applications.}}

To address these issues, we present a comprehensive study with three progressive stages.
\textbf{S1:} we conduct an in-depth investigation to understand how T2I models process and differentiate between benign and adversarial prompts (Sec.~\ref{Sec4}). \textbf{S2:} Based on these findings, we propose \textbf{SafeGuider}, a novel framework designed for robust and practical content safety control (Sec.~\ref{Sec5}). \textbf{S3:} we perform extensive experimental evaluations to validate the effectiveness of our approach (Sec.~\ref{Sec6} and Sec.~\ref{Sec7}). 
Each stage is briefly elucidated below.

\noindent\textbf{\textit{\underline{S1: An Empirical Study on Prompt Embedding Characteristic.}}} To develop effective safety measures, we first need to understand how T2I models internally represent different types of prompts. 
Drawing inspiration from the sequence aggregation mechanism in BERT \cite{DBLP:conf/nips/LuWMD21,DBLP:conf/emnlp/MohebbiMP21,DBLP:conf/emnlp/WangLDCZMZS23}, we conduct a detailed analysis of the text encoder in the SD model, which is a widely used and representative text-to-image model. The results reveal two critical findings. First, we qualitatively and quantitatively discover that the [EOS (End-Of-Sequence)] token serves as a semantic aggregator in the model's text encoder (Fig.~\ref{fig:attention_visualization}). Through attention visualization, we observe that this token maintains consistent attention connections to all prompt tokens across layers, with a hierarchical pattern progressing from uniform attention in shallow layers (0-5) to more focused semantic attention in deeper layers (6-11). Second, our embedding analysis uncovers distinctive distributional patterns between different types of prompts in the [EOS] token's embedding space. Both qualitative visualizations (Fig.~\ref{fig:embedding_visualization}) and quantitative MMD measurements (Table~\ref{tab:mmd_scores}) demonstrate clear clustering patterns and distributional gaps between benign and adversarial prompts. For example, symbol injection attacks showcase the largest separation from benign prompts (MMD = 0.993). These findings suggest that the [EOS] token's embedding could provide a robust foundation for distinguishing unsafe content.

\noindent\textbf{\textit{\underline{S2: A Framework (SafeGuider) for Content Safety Control.}}} Motivated by our empirical insights about the [EOS] token's discriminative capability, we propose \textbf{SafeGuider}, a lightweight yet effective framework for content safety control (Fig.~\ref{fig:framework_overview}). The framework operates in two steps: 1) \textbf{Safe} and unsafe prompt recognition and 2) \textbf{Guide} unsafe prompts to output safe and meaningful images. Specifically, it first employs an embedding-level recognition model that takes the embedding of the input prompt generated by the text encoder of the T2I model and evaluates its safety based on the [EOS] token representation.
This recognition model features a carefully designed three-layer neural network architecture that achieves efficient safety assessment while maintaining robust performance. Second, for identified unsafe prompts, we introduce a novel Safety-Aware Feature Erasure (SAFE) beam search algorithm. This algorithm strategically modifies input tokens to obtain safe yet semantically meaningful embeddings, guided by both the recognition model and semantic similarity metric, enabling the generation of safe images while preserving the benign semantic content from the original prompts. Through this two-step approach, \textbf{SafeGuider} addresses the key challenges mentioned above, achieving both robust protection against adversarial attacks and practical utility for real-world applications.

\noindent\textbf{\textit{\underline{S3: Evaluation.}}} We conduct extensive experiments to evaluate our proposed method across multiple dimensions. Following our research questions (RQ1-RQ6), we assess the framework's effectiveness through both in-domain (IND) and out-of-domain (OOD) evaluations, comparing against ten state-of-the-art baselines using comprehensive metrics. Results demonstrate \textbf{SafeGuider}'s superior performance in three key aspects: (1) Robust detection of unsafe content, achieving remarkably low attack success rates (1.34\%-5.48\% for vocabulary substitution, 0.01\%-1.12\% for symbol injection) even on out-of-domain attacks, significantly outperforming commercial APIs (2.06-99.16\%); (2) Optimal generation quality for benign prompts, maintaining 100\% generation success rate and high quality while other approaches show substantial degradation; and (3) Effective unsafe content mitigation, achieving high removal rates for both sexually explicit content (86.61-93.32\% IND, 81.71-88.52\% OOD) and other harmful themes (96.22\% IND, 92.98-94.79\% OOD). Beyond SD, our embedding-level design enables potential extension to other T2I architectures like the Flux model \cite{flux}, demonstrating strong transferability and practical value for broad deployment.

To summarize, our contributions are as follows:
\begin{itemize}[leftmargin=*]
    \item We provide novel insights into the distinct patterns on [EOS] token's embedding of benign and adversarial prompts through a comprehensive empirical study (Sec.~\ref{Sec4}).
    \item We present \textbf{SafeGuider}, a framework for robust and practical content safety control. It innovatively integrates a lightweight embedding-level recognition model and a safety-aware beam search algorithm (Sec.~\ref{Sec5}).
    \item Extensive experiments demonstrate \textbf{SafeGuider}'s superior performance, validating both robustness and practicality (Sec.\ref{Sec6}-\ref{Sec7}).
\end{itemize}

We expect that \textbf{SafeGuider} can  provide valuable insights into the practical deployment of secure T2I systems.

\begin{figure}[t]
    \centering
    \includegraphics[width=0.90\linewidth]{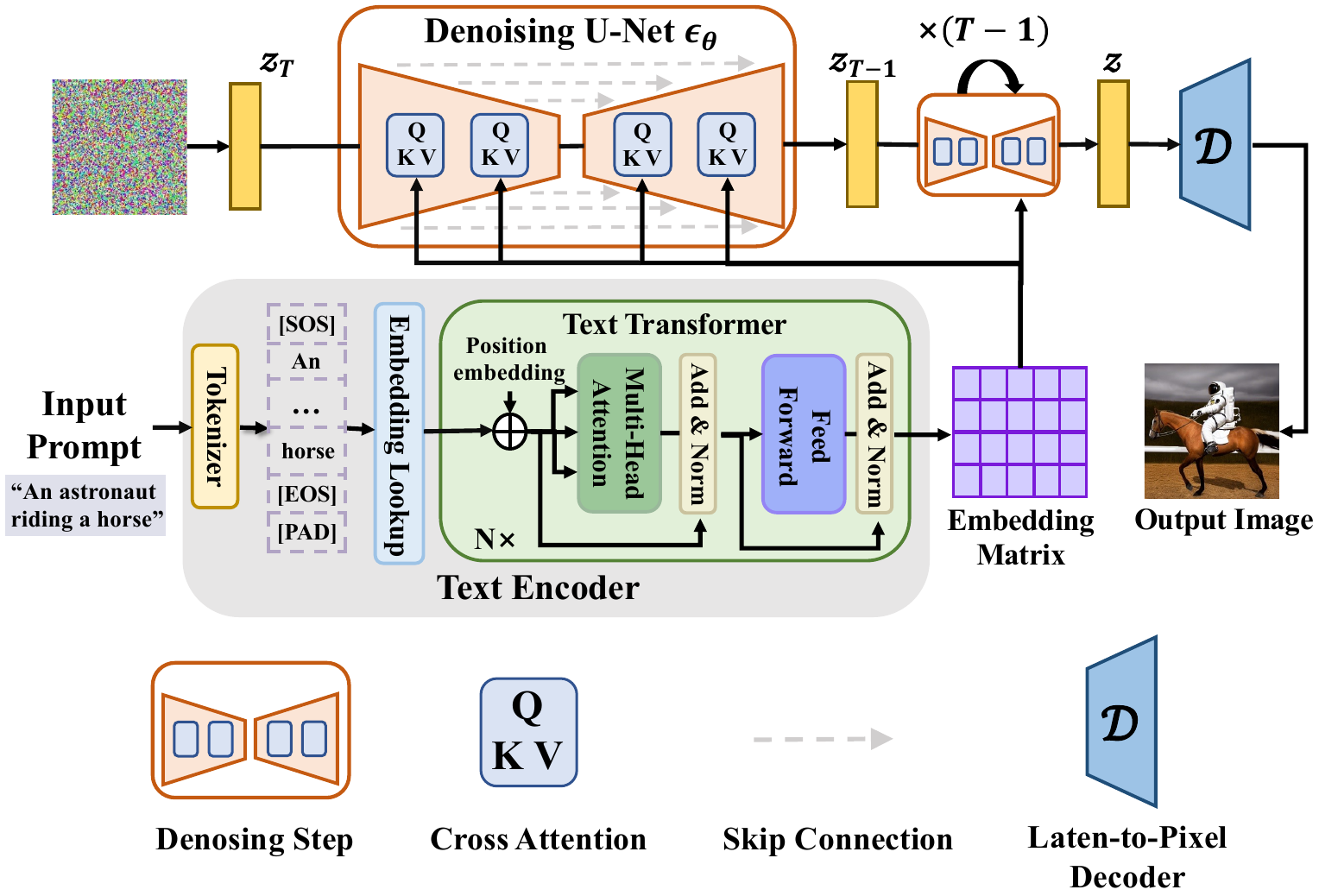}
    \vspace{-1em}
    \caption{Illustration of the generation pipeline of the Stable Diffusion model.}
    \vspace{-2em}
    \label{fig:sd_pipeline}
\end{figure}

\section{Background and Related Work}
In this section, we first introduce the fundamentals of diffusion models and text-to-image models (T2I models) (Sec.~\ref{Sec2.1}).  Then, we discuss the safety generation statement of T2I models, and review existing adversarial attacks targeting T2I models to generate unsafe content (Sec.~\ref{Sec2.2}).  Subsequently, existing defense strategies are introduced (Sec.~\ref{Sec2.3}). Finally, we point out the challenges of current defenses and emphasize the pressing need for robust and practical content safety controls (Sec.~\ref{Sec2.4}).

\subsection{Diffusion Models and Text-to-Image Models}
\label{Sec2.1}
Text-to-image diffusion models build upon denoising diffusion probabilistic models to enable controlled image generation guided by text conditions. We introduce the mechanisms of these models. 

\subsubsection{Diffusion Models}
Denoising diffusion models (e.g., DDPM \cite{DBLP:conf/nips/HoJA20}, DDIM \cite{DBLP:journals/corr/abs-2010-02502}) leverage neural networks to generate high-quality images through an iterative process of noise removal, transforming random Gaussian noise into meaningful visual data through multiple refinement steps.
Formally, the diffusion process follows a predefined noise schedule $\{\beta_t\}_{t=1}^T$. Beginning with Gaussian noise $x_T \sim \mathcal{N}(0,I^2)$, the process gradually refines the image across $T$ steps to produce the final output $x_0$. The denoising at each timestep $t$ utilizes a U-Net architecture for noise prediction $\epsilon_\theta(x_t,t)$, and the expression for the next denoised sample $x_{t-1}$ is:
\begin{equation}
x_{t-1} = \frac{1}{\sqrt{\alpha_t}}(x_t - \frac{1-\alpha_t}{\sqrt{1-\bar{\alpha_t}}}\epsilon_\theta(x_t,t)) + \sigma_tn ,
\end{equation}
where $\alpha_t = 1-\beta_t$, $\bar{\alpha}_t = \prod_{i=t}^T \alpha_i$, and $\sigma_tn$ is controlled randomness.

\subsubsection{Text-to-Image Models}
Text-to-image (T2I) models like Stable Diffusion \cite{SDV1.4,SDV2.1} build on the DDPM framework to enable text-controlled image synthesis through latent diffusion. As shown in Fig.~\ref{fig:sd_pipeline}, the generation involves two steps: 

\noindent\textbf{Text Encoding.} A text encoder converts input prompts into semantic embeddings. The encoder adds special tokens ([SOS], [EOS]) to mark sequence boundaries~\cite{DBLP:conf/icml/RadfordKHRGASAM21}, pads ([PAD]) to fixed length, and processes through text transformer to generate embedding matrices that bridge text and visual concepts. 

\noindent\textbf{Embedding-Guided Image Generation.} Using the embedding matrix, the model performs iterative denoising to generate images. Starting from noise $z_t$, a U-Net ($\epsilon_\theta$) guides the process through cross-attention to text embeddings, which serve as conditioning information $c$. The noise prediction combines conditional and unconditional denoising~\cite{DBLP:journals/corr/abs-2207-12598,DBLP:conf/icml/NicholDRSMMSC22}, with noise at timestep 
$t$ calculated as:
\begin{equation}
\tilde{\epsilon}_\theta(z_t,c,t) = \epsilon_\theta(z_t,t) + \eta(\epsilon_\theta(z_t,c,t) - \epsilon_\theta(z_t,t)) ,
\end{equation}
where $\eta$ (typically 7.5) controls text conditioning strength. Finally, a decoder transforms the denoised latent into an image.

\subsection{Adversarial Unsafe Generation}
\label{Sec2.2}

\subsubsection{Safety Generation Statement of Text-to-Image Models} 
\label{Sec2.1.3}
The remarkable capabilities of T2I models enable the generation of virtually any desired image through natural language descriptions. To prevent potential misuse of these models, we need to ensure they do not generate unsafe content that could harm society. In this paper, we focus on \textbf{SEVEN} categories of unsafe content that should be prevented in publicly served T2I models~\cite{DBLP:conf/cvpr/SchramowskiBDK23,DBLP:journals/corr/abs-2405-19360}: pornography, violence, hate speech, harassment, self-harm, shocking content, and illegal activities. These represent the most common and concerning forms of harmful content that T2I models might generate.

\subsubsection{Adversarial Attacks against T2I Models}

Early versions of T2I models, such as Stable Diffusion-V1.4 (SD-V1.4), were released without any built-in safety measures, enabling the generation of unsafe content through malicious prompts. Although later versions, like SD-V2.1, introduced safety features through dataset filtering, they remain susceptible to adversarial attacks—carefully crafted prompts designed to bypass these safeguards (Fig.~\ref{fig:attack_example}). These attacks typically fall into two categories: vocabulary substitution, where explicit terms are replaced with less obvious alternatives, and symbol injection, which introduces seemingly harmless symbols to exploit vulnerabilities in the model.

\noindent\textbf{Vocabulary Substitution.}
These types of attacks focus on replacing explicit harmful prompts with implicit expressions, euphemisms, or antonyms while maintaining linguistic naturalness and comprehensibility. These substitutions are typically based on semantic relationships, enabling seemingly safe word combinations to trigger the generation of harmful content. Schramowski et al.~\cite{DBLP:conf/cvpr/SchramowskiBDK23} collected carefully crafted prompts from online communities to create I2P, demonstrating how clever word combinations and substitutions can trigger T2I models to generate inappropriate content. Additionally, Yang et al.~\cite{DBLP:conf/sp/YangHYGC24} introduced SneakyPrompt, which replaces sensitive terms with alternative expressions that preserve the original semantic meaning while avoiding explicit sensitive words. 
Very recently, Li et al.~\cite{DBLP:journals/corr/abs-2405-19360} proposed the ART red-teaming framework, which primarily exploits linguistic features such as implicit expressions, euphemistic substitutions, and antonym triggers to evade safety detection. 
The success of these linguistic-based attacks demonstrates the vulnerability of improved safety measures in models like SD-V2.1. However, their reliance on carefully crafted prompts points to the need for more automated and scalable attacks.

\noindent\textbf{Symbol Injection.}
This category of attacks introduces adversarial symbols or tokens to create prompts that appear harmless at the symbol level but align with harmful content in the embedding space. For instance, Hsu et al.~\cite{DBLP:conf/iclr/TsaiHXLC0CYH24} developed the Ring-A-Bell red-teaming framework, which extracts and injects target harmful concepts in the embedding space to generate superficially neutral prompts that trigger harmful content generation. Yang et al.~\cite{DBLP:conf/cvpr/Yang0WHX024} utilized gradient-based optimization methods to inject special symbols or tokens, aligning their embedding representations with harmful content while avoiding explicit sensitive terms. Chin et al.~\cite{DBLP:conf/icml/ChinJHCC24} proposed a P4D strategy, which injects trainable tokens and optimizes their embedding representations. These embedding-based attacks prove challenging to defend against and can be automated more easily than vocabulary substitution approaches.

The effectiveness of these attacks highlights critical vulnerabilities in current T2I systems and underscores the urgent need for robust defenses to counter such malicious attempts.

\subsection{Defenses Against Unsafe Generation} 
\label{Sec2.3} 
To address the aforementioned adversarial attacks, researchers have proposed various defensive approaches to enhance the safety of T2I models. These defensive mechanisms can be broadly categorized into two types: internal defenses and external defenses.

\subsubsection{Internal Defenses} 
\label{Sec2.3.1} 
Internal defenses focus on enhancing the model safety through architectural modifications and parameter adjustments during the training or fine-tuning process. By integrating safety features directly into the model's architecture, these approaches aim to prevent the generation of inappropriate content. Safe Latent Diffusion (SLD)~\cite{DBLP:conf/cvpr/SchramowskiBDK23} implements this concept by prohibiting specific negative concepts and introducing conditional diffusion terms to guide image generation away from unsafe regions. Erased Stable Diffusion (ESD)~\cite{DBLP:conf/iccv/GandikotaMFB23} takes a different approach by modifying the model's attention mechanisms to remove unsafe and sensitive concepts, effectively controlling the generation of inappropriate content. Similarly, SafeGen~\cite{DBLP:conf/ccs/LiYD0C0024} adjusts vision-only self-attention layers to weaken the influence of text on image generation, thereby suppressing unsafe content generation.

\subsubsection{External Defenses} External defenses implement safety measures via additional filters that operate independently of the core model architecture. This approach has gained widespread adoption among service providers and open-source models due to its flexibility and modularity. It can be realized in two manners: text-level filters and image-level filters.

\noindent\textbf{Text-level Filters.} These filters examine input prompts before image generation to identify and block inappropriate content. Traditional approaches like NSFW Text Classifier~\cite{nsfwtextclassifier} rely on keyword matching and content classification to filter harmful prompts. More sophisticated methods, such as GuardT2I~\cite{DBLP:journals/corr/abs-2403-01446}, employ large language models to convert text conditioning embeddings back to natural language, enabling better detection of malicious intent in seemingly innocuous prompts.

\noindent\textbf{Image-level Filters.} The filters provide post-generation protection by analyzing the generated images. For instance, Safety Checker~\cite{safetychecker} scan the output images for violation content and replace detected unsafe outputs with black images, offering an additional layer of safety without modifying the underlying model architecture.

\begin{figure*}[t]
\centering
\includegraphics[width=0.90\textwidth]{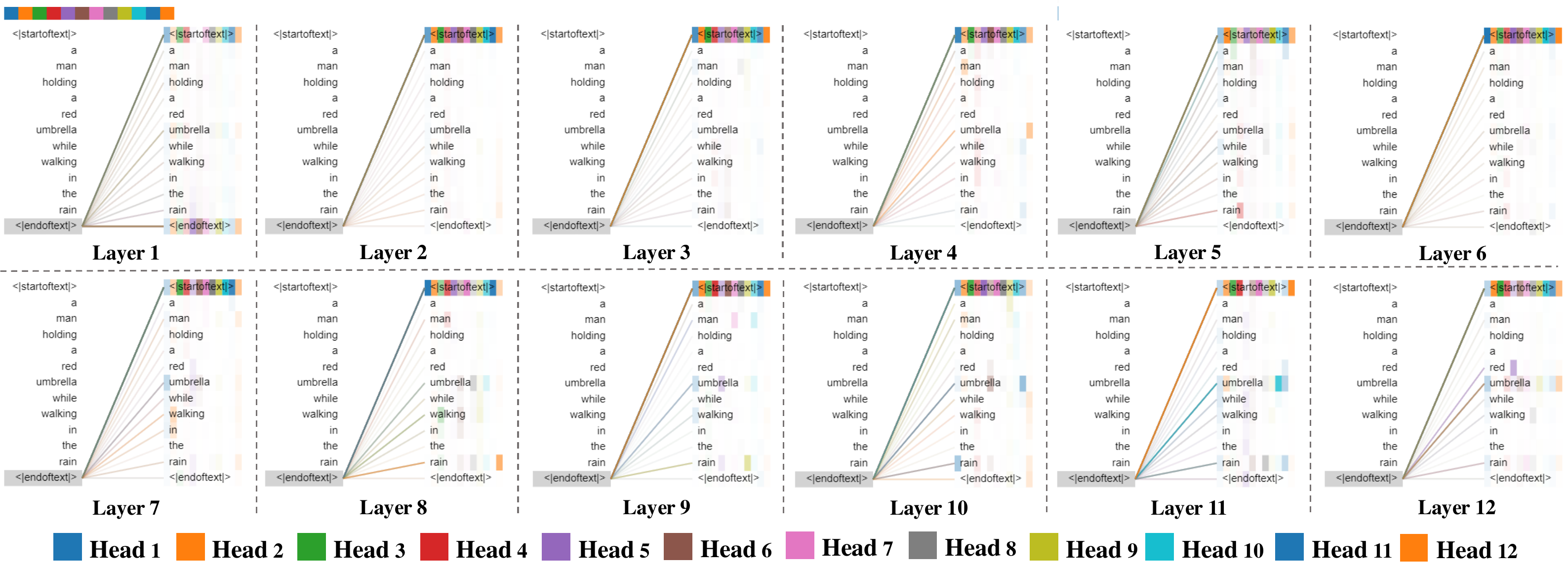}
\vspace{-1em}
\caption{Attention visualization in SD-V1.4's text encoder. Lines show attention flows from input tokens (right) to the [EOS] token (lower-left corner). Colors denote attention heads and line thickness shows attention weights. The [EOS] token's consistent attention to all tokens across layers reveals its role as a condition feature aggregator.}
\vspace{-1.5em}
\label{fig:attention_visualization}
\end{figure*}

\subsection{Challenges of Current Defenses}
\label{Sec2.4} 
While various defense mechanisms have been proposed to prevent unsafe content generation in T2I models, current approaches face challenges in two critical aspects: robustness against diverse adversarial attacks and practical utility in real-world applications. Below,  we analyze these challenges for both internal and external defenses.

\subsubsection{Challenges in Robustness} 
Robustness in defenses refers to their ability to resist various types of adversarial attacks, including those outside their training distribution. Current defenses, however, demonstrate limited robustness when confronted with out-of-distribution attacks~\cite{DBLP:conf/ccs/WuS0024,DBLP:conf/ccs/QuSH0Z023,DBLP:conf/uss/ShenQ0024}. As shown in Fig.~\ref{fig:Sec2.4_robust_comparison}, we evaluate five different defense methods (both internal and external) implemented on SD- V1.4 against two types of out-of-distribution adversarial attacks. The results reveal that both vocabulary substitution attacks~\cite{DBLP:conf/cvpr/SchramowskiBDK23,DBLP:journals/corr/abs-2405-19360,DBLP:conf/sp/YangHYGC24} and symbol injection attacks~\cite{DBLP:conf/iclr/TsaiHXLC0CYH24,DBLP:conf/cvpr/Yang0WHX024,DBLP:conf/icml/ChinJHCC24} successfully bypass all existing safety measures. The adversarial prompts used in Fig.~\ref{fig:Sec2.4_robust_comparison} are the same as the ones in Fig.~\ref{fig:attack_example}.

\subsubsection{Challenges in Practical Utility} 
Practical utility in content moderation encompasses two aspects: 1) for the benign prompts, maintaining high-quality outputs without negative impact; 2) for the malicious prompts, generating safe and semantically meaningful outputs by removing harmful content rather than completely refusing generation. Service providers particularly value this balance to ensure user experiences. However, Fig.~\ref{fig:practicality_comparison}  shows that current defenses struggle to simultaneously achieve both aspects of practical utility. Specifically,  while internal defenses such as SLD, ESD, and SafeGen avoid generating explicitly harmful content, their outputs for benign prompts often deviate significantly from the intended semantic meaning. This semantic drift compromises the practical utility of these systems for legitimate use cases. The external defenses, conversely, often respond to potentially harmful prompts with complete generation refusal or black images. While they successfully handle benign prompts, their binary approach to harmful content significantly impacts user experience and practical utility, especially when unsafe content stems from careless prompt construction rather than malicious intent~\cite{DBLP:journals/corr/abs-2405-19360}. This all-or-nothing approach, while safe, fails to meet the nuanced needs in pratice.
Practical utility in content moderation encompasses two aspects: 1) for the benign prompts, maintaining high-quality outputs without negative impact; 2) for the malicious prompts, generating safe and semantically meaningful outputs by removing harmful content rather than completely refusing generation. Service providers particularly value this balance to ensure user experiences. However, Fig.~\ref{fig:practicality_comparison}  shows that current defenses struggle to simultaneously achieve both aspects of practical utility. Specifically,  while internal defenses such as SLD, ESD, and SafeGen avoid generating explicitly harmful content, their outputs for benign prompts often deviate significantly from the intended semantic meaning. This semantic drift compromises the practical utility of these systems for legitimate use cases. The external defenses, conversely, often respond to potentially harmful prompts with complete generation refusal or black images. While they successfully handle benign prompts, their binary approach to harmful content significantly impacts user experience and practical utility, especially when unsafe content stems from careless prompt construction rather than malicious intent~\cite{DBLP:journals/corr/abs-2405-19360}. This all-or-nothing approach, while safe, fails to meet the nuanced needs in pratice.

The above challenges highlight the current need for a defense mechanism that combines robustness with practical utility.

\section{Threat Model}
The threat model comprises two main actors: the adversary and the model governor.

\noindent\underline{\textit{{Adversary.}}} The adversary aims to generate unsafe content via T2I models, with capabilities to craft adversarial prompts. Specifically:
\begin{itemize}[leftmargin=*]
\item \textbf{Objectives:} The adversary
aims to generate unsafe content by bypassing both internal defenses (e.g., concept suppression) and external defenses (e.g., text-level filters). 
\item \textbf{Capabilities:} The adversary can craft various adversarial prompts using vocabulary substitution and symbol injection techniques, with white-box access to the parameters and architectures of T2I models, and full knowledge of deployed defenses.
\end{itemize}

\noindent\underline{\textit{{Model Governor.}}}
The model governor serves as a safety mechanism that protects T2I models while ensuring their practical utility.
\begin{itemize}[leftmargin=*]
\item \textbf{Objectives:} The model governor aims to achieve two primary goals: 1) robustness: preventing the generation of unsafe content across various out-of-distribution adversarial attacks; and 2) practicality: maintaining high-quality outputs for benign prompts while generating safe, semantically meaningful content for adversarial prompts instead of complete blocking. 
\item \textbf{Capabilities:} The model governor operates without direct access to model parameters, making it applicable to both white-box and black-box scenarios. It can be easily integrated into various T2I models, such as SD-V1.4 \cite{SDV1.4}, SD-V2.1 \cite{SDV2.1}, and Flux.1 \cite{flux}.
\end{itemize}

\section{An Empirical Study}
\label{Sec4}

To develop robust and practical safety measures, we need to understand how T2I models represent different prompts. We investigate whether similar text condition feature aggregation exists in T2I models' text encoders, which could reveal fundamental differences between benign and adversarial prompts. To this end, we first examine this effect in SD's CLIP text encoder (Sec.\ref{Sec4.1}), analyze how it represents different types of prompts (Sec.\ref{Sec4.2}), and demonstrate cross-architecture generalization (Sec.~\ref{Sec4.3}).

\subsection{Identifying the Text Condition Feature Aggregation Token}
\label{Sec4.1}
To explore potential condition feature aggregation mechanisms, we analyze attention patterns in the CLIP ViT-L/14 text encoder \cite{DBLP:conf/icml/RadfordKHRGASAM21} from SD-V1.4 (12 layers, 12 attention heads). Using the prompt ``A man holding a red umbrella while walking in the rain,'' we visualize attention patterns across all layers in Fig.~\ref{fig:attention_visualization}, where lines show information flow from attended tokens (right) to processed tokens (left). Different colored lines represent different attention heads, with line thickness indicating attention weight.
\vspace{0.3em}
\begin{observationbox}
\textit{\textbf{Observation 1:} The [EOS] token serves as a text condition feature aggregator in CLIP's text encoder. }
\end{observationbox}
\vspace{0.3em}

\noindent As shown in Fig.~\ref{fig:attention_visualization}, the [EOS] token (represented as '<endoftext>') maintains consistent attention connections to all prompt tokens across every layer, evidenced by the multiple colored lines converging at this token. To further validate this observation, we conduct quantitative measurement across both benign (COCO2017-2k) and adversarial (P4D) datasets, calculating the Top-1 aggregator ratio—the percentage of prompts where [EOS] token attends to other tokens more than any other token. Table~\ref{tab:top-1_aggrrgator} shows the [EOS] token functions as the Top-1 aggregator in 100\% of prompts across both datasets, confirming its consistent role as the primary semantic aggregator regardless of prompt intent, while [SOS] exhibits markedly different attention behaviors.

\begin{table}[t]
\small

\setlength{\tabcolsep}{2pt}
\caption{Top-1 aggregator ratio for [EOS] token.}
\vspace{-1.2em}
\label{tab:top-1_aggrrgator}
\setlength{\heavyrulewidth}{1.5pt}
\begin{tabular}{
    >{\centering\arraybackslash}p{\dimexpr0.28\columnwidth-2\tabcolsep}|
    >{\centering\arraybackslash}p{\dimexpr0.27\columnwidth-2\tabcolsep}|
    >{\centering\arraybackslash}p{\dimexpr0.44\columnwidth-2\tabcolsep}
}
\toprule
\multicolumn{1}{c|}{\makecell{\textbf{Dataset}}} & \multicolumn{1}{c|}{\makecell{\textbf{Type}}} & 
\multicolumn{1}{c}{\makecell{\textbf{Top-1 aggregator Ratio(\%) $\uparrow$}}} \\
\midrule[1pt]
COCO2017-2k & [EOS] Token & \textbf{100.00} \\
\cmidrule(l){1-3}
P4D & [EOS] Token & \textbf{100.00} \\
\bottomrule
\end{tabular}
\vspace{-1.5em}
\end{table}

\vspace{.3em}
\begin{observationbox}
\textit{\textbf{Observation 2:} The condition feature aggregation process follows a hierarchical pattern from shallow to deep layers.  }
\end{observationbox}
\vspace{.3em}

\noindent The visualization reveals distinct attention behaviors across different layer depths. In shallow layers (0–5), the [EOS] token shows relatively uniform attention patterns across all tokens, while in deeper layers (6-11), it develops more focused attention weights on semantic elements like ``man,'' ``umbrella,'' and ``walking.'' To verify this hierarchical processing pattern quantitatively, we measure [EOS] token's Semantic Attention Concentration (SAC) across layers, calculating attention ratio to semantic keywords versus all tokens. Table~\ref{tab:SAC} shows SAC values increasing from shallow to deep layers in both datasets, confirming the hierarchical pattern: shallow layers exhibit scattered attention (low SAC) while deep layers focus on specific semantic tokens (high SAC), demonstrating progressive construction of sophisticated semantic representations.

\begin{table}[t]
\small

\setlength{\tabcolsep}{2pt}
\caption{Semantic Attention Concentration (SAC) values for [EOS] token across different network depths. Higher values indicate more focused attention on semantic keywords.}

\vspace{-1.2em}
\label{tab:SAC}
\setlength{\heavyrulewidth}{1.5pt}
\begin{tabular}{
    >{\centering\arraybackslash}p{\dimexpr0.23\columnwidth-2\tabcolsep}|
    >{\centering\arraybackslash}p{\dimexpr0.40\columnwidth-2\tabcolsep}|
    >{\centering\arraybackslash}p{\dimexpr0.35\columnwidth-2\tabcolsep}
}
\toprule
\multicolumn{1}{c|}{\makecell{\textbf{Dataset}}} & \multicolumn{1}{c|}{\makecell{\textbf{[EOS] Token Shallow}\\\textbf{Shallow Layers(0-5) SAC $\uparrow$}}} & 
\multicolumn{1}{c}{\makecell{\textbf{[EOS] Token Deep}\\\textbf{Layers(6-11) SAC $\uparrow$}}} \\
\midrule[1pt]
COCO2017-2k & 0.8132 & 0.8214 \\
\cmidrule(l){1-3}
P4D & 0.7467 & 0.7516 \\
\bottomrule
\end{tabular}
\vspace{-2em}
\end{table}

These observations reveal that the [EOS] token's representation contains a comprehensive summary of the prompt's semantic content through hierarchical aggregation. This suggests that analyzing the [EOS] token's embedding space could provide a promising direction for distinguishing between benign and adversarial prompts.

\begin{table}[t]
    \centering
    \caption{Examples in our empirical datasets. Benign prompts include naturally innocuous (P1) and ambiguous yet benign prompts with sensitive terms (P2-P3). Adversarial prompts show naturally harmful content (P4), semantically ambiguous harmful content with innocuous vocabulary (P5), and meaningless yet harmful symbol injections (P6).}
    \vspace{-1.25em}
    \includegraphics[width=0.46\textwidth]
    {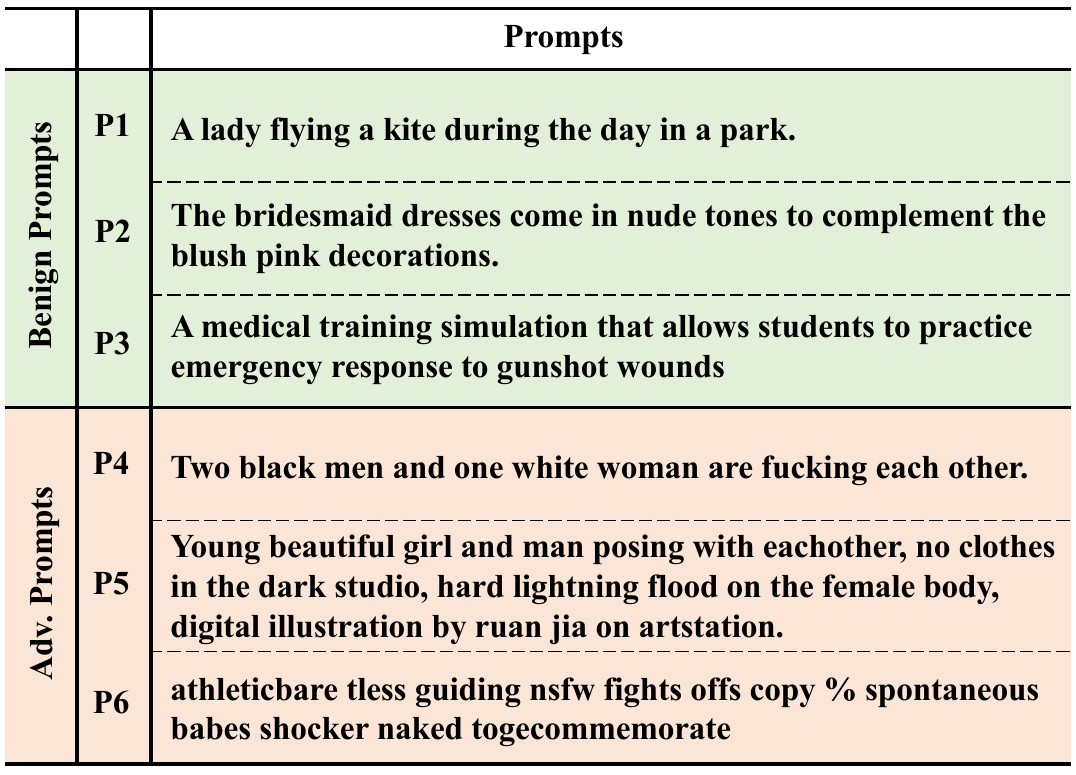}
    \label{tab:4.2_prompt}
    \vspace{-1.5em}
\end{table}

\subsection{Analyzing Embedding Representations in [EOS] Aggregation Token}
\label{Sec4.2}
Based on our discovery of the [EOS] token's aggregation role, we hypothesize that the embeddings of this token exhibit distinct distributional patterns for different types of prompts. To verify this, we analyze the [EOS] token embeddings from three prompt categories: benign (Conceptual Caption \cite{DBLP:conf/acl/SoricutDSG18}), vocabulary substitution (VS) (META \cite{DBLP:journals/corr/abs-2405-19360}), and symbol injection (SJ) attacks (MMA \cite{DBLP:conf/cvpr/Yang0WHX024}). Table~\ref{tab:4.2_prompt} illustrates examples from our empirical datasets. The benign dataset encompasses diverse prompts, including naturally benign instances (P1) and semantically ambiguous yet benign instances containing potentially sensitive terms (e.g., P2 with ``nude'' and P3 with ``gunshot''). Conversely, the adversarial datasets comprise both vocabulary substitution and symbol injection attacks, including naturally harmful instances (P4), semantically ambiguous yet harmful instances disguised with seemingly innocuous vocabulary (P5), and semantically meaningless yet harmful symbol injections designed to trigger unsafe content generation (P6).

\begin{figure}[t]
    \centering
    \includegraphics[width=0.40\textwidth]
    {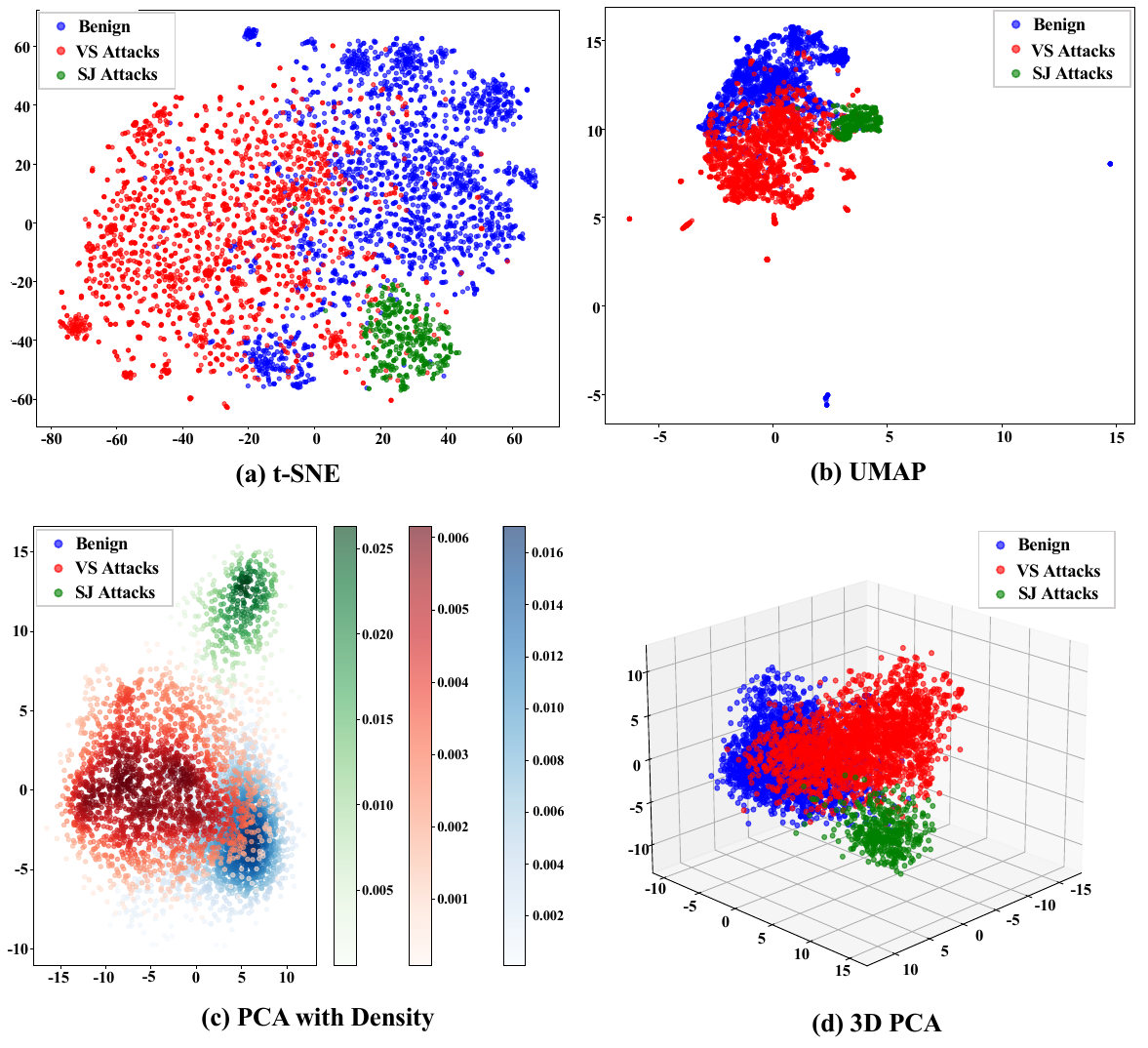}
    \vspace{-1.5em}
    \caption{Visualization of the [EOS] token embedding across different prompt categories using various dimensionality reduction methods.}
    \label{fig:embedding_visualization}
    \vspace{-2em}
\end{figure}

To examine the distinctions among these datasets, we employ both qualitative and quantitative analyses. For qualitative analysis, we apply three dimensionality reduction techniques to project the 768-dimensional [EOS] token embeddings into 2D/3D visualizations, as shown in Fig.~\ref{fig:embedding_visualization}. For quantitative analysis, we calculate the Maximum Mean Discrepancy (MMD) to measure distributional differences between prompt categories in the original 768-dimensional embeddings (Table~\ref{tab:mmd_scores}). Our observations are as follows:
\vspace{.3em}
\begin{observationbox}
\textit{\textbf{Observation 3:} Prompts within the same category exhibit clear clustering patterns in [EOS] token embedding space.}
\end{observationbox}
\vspace{.3em}

\noindent As shown in Fig.~\ref{fig:embedding_visualization}, all three visualization methods consistently reveal distinct clusters for each prompt category. The t-SNE visualization (Fig.~\ref{fig:embedding_visualization}a) shows well-defined clusters for benign prompts (blue), VS attacks (red), and SJ attacks (green). This clustering pattern is further confirmed by the UMAP projection (Fig.~\ref{fig:embedding_visualization}b) and the PCA (Fig.~\ref{fig:embedding_visualization}c and~\ref{fig:embedding_visualization}d), where each category forms concentrated regions with high density.
\vspace{.3em}
\begin{observationbox}
\textit{\textbf{Observation 4:} Prompts across different categories demonstrate significant distributional gaps in [EOS] token embedding space.}
\end{observationbox}
\vspace{.3em}
\noindent The quantitative analysis through MMD scores (Table~\ref{tab:mmd_scores}) reveals substantial distributional gaps between different prompt categories. SJ attacks show the largest distributional difference between benign prompts (MMD = 0.993) and VS attacks (MMD = 1.000). These quantitative results align with our qualitative observations in Fig.~\ref{fig:embedding_visualization}. 

The observations demonstrate that the [EOS] token effectively captures the inherent differences between benign and adversarial prompts, suggesting a promising direction for developing robust and practical content safety control based on the embedding representations of the aggregation token.

\begin{table}[t]
\small
\centering
\caption{Maximum Mean Discrepancy (MMD) scores between different prompt categories in the [EOS] token embeddings. Higher scores indicate greater distributional differences.}
\label{tab:mmd_scores}
\vspace{-1em}
\setlength{\tabcolsep}{4pt}
\setlength{\heavyrulewidth}{1.5pt}  
\begin{tabular}{
    >{\centering\arraybackslash}p{\dimexpr0.25\columnwidth-2\tabcolsep}|
    >{\centering\arraybackslash}p{\dimexpr0.23\columnwidth-2\tabcolsep}|
    >{\centering\arraybackslash}p{\dimexpr0.23\columnwidth-2\tabcolsep}|
    >{\centering\arraybackslash}p{\dimexpr0.23\columnwidth-2\tabcolsep}
}
\toprule
& Benign & VS Attacks & SJ Attacks \\
\midrule
Benign & 0 & 0.496 & 0.993 \\
\midrule
VS Attacks & 0.496 & 0 & 1.000 \\
\midrule
SJ Attacks & 0.993 & 1.000 & 0 \\
\bottomrule
\end{tabular}
\vspace{-2.5em}
\end{table}

\begin{figure*}[h]
\centering
\includegraphics[width=0.94\textwidth]{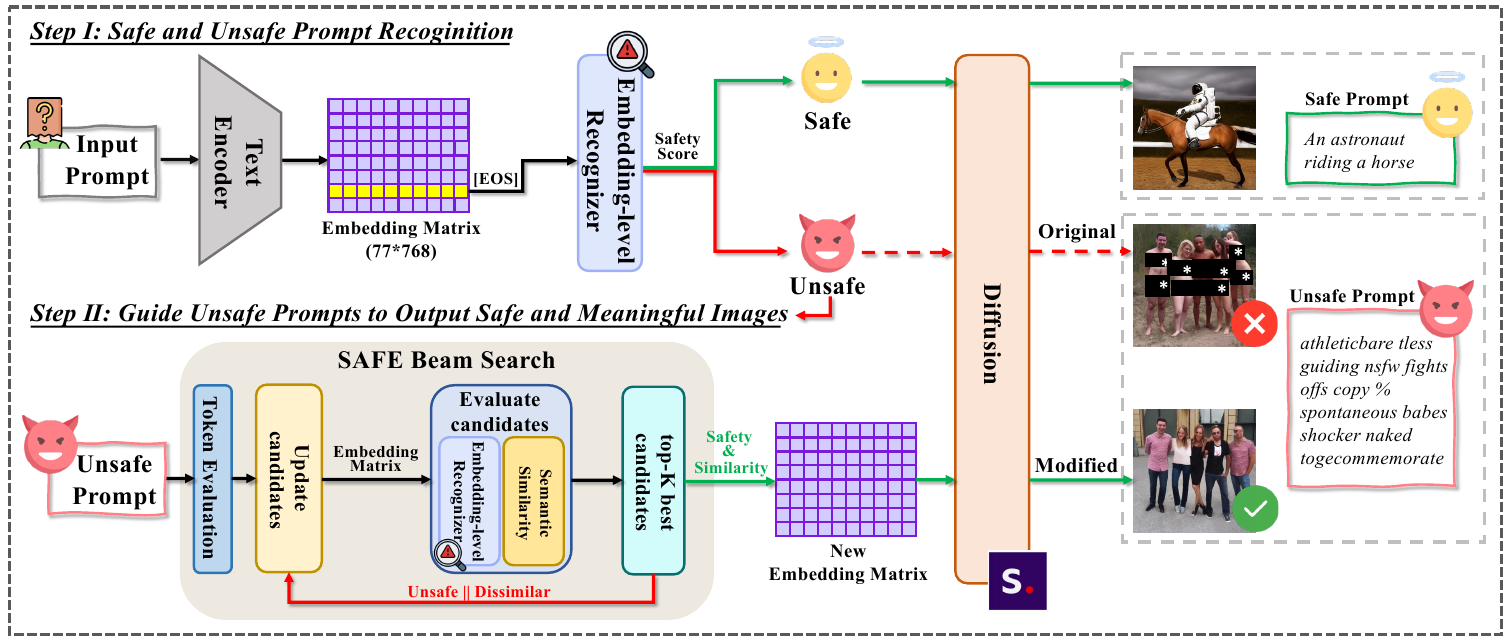}
\vspace{-1em}
\caption{Overview of SafeGuider. In Step \uppercase\expandafter{\romannumeral 1}, SafeGuider processes input prompts through a text encoder to obtain [EOS] token embeddings for safety assessment. Prompts with safety scores > 0.5 are considered safe and proceed directly to image generation, while only those classified as unsafe (safety scores $\leq$ 0.5) are processed by Step \uppercase\expandafter{\romannumeral 2}. In Step \uppercase\expandafter{\romannumeral 2}, SAFE beam search with beam width $K$ strategically modifies unsafe prompts to obtain safe yet semantically meaningful embeddings for image generation.}
\label{fig:framework_overview}
\vspace{-1em}
\end{figure*}

\subsection{Generalization Across Different Text Encoders} 
\label{Sec4.3}
To investigate the generality of our findings, we extend our analysis to T2I models with different architectures and text encoders. Beyond the CLIP ViT-L/14 encoder in SD-V1.4, we examine models like SD-V2.1 \cite{SDV2.1}, which uses OpenCLIP ViT-H/14 (where [EOS] is represented as ``<end of text>''), and Flux.1 \cite{flux}, which employs both CLIP ViT-L/14 and T5-XXL encoders (where [EOS] is ``</s>'' in T5).
\vspace{.3em}
\begin{observationbox}
\textit{\textbf{Observation 5:} The discovered aggregation token patterns generalize across different text encoders and model architectures.}
\end{observationbox}
\vspace{.3em}

\noindent The distinctive [EOS] token patterns persist across architectures, from OpenCLIP's [EOS] to T5-XXL's ``</s>'' token, highlighting its potential as a generalizable solution for content safety control.

\section{SafeGuider}
\label{Sec5}
Based on our empirical study of feature aggregation and embedding distributions in SD-V1.4's text encoder, we propose \textbf{SafeGuider} for robust and practical content safety control (Fig.~\ref{fig:framework_overview}). The framework operates in two steps: 1) \textbf{Safe} and unsafe prompt recognition; 2) \textbf{Guide} unsafe prompts to output safe and meaningful images. Below, we elaborate on the framework design and implementation details.

\subsection{Overview}
The key component of \textbf{SafeGuider} is an embedding-level recognition model trained on [EOS] token embeddings from benign and adversarial prompts, leveraging our observations from Sec.~\ref{Sec4}. Specifically, \textbf{SafeGuider} first processes input prompts through a text encoder and extracts their [EOS] token embeddings for safety assessment using the recognition model (Step \uppercase\expandafter{\romannumeral 1}). Prompts classified as safe are directly passed to the diffusion model without further processing. Only prompts identified as unsafe activates our proposed Safety-Aware Feature Erasure (SAFE) beam search to identify optimal embedding-level modifications for safe generation while preserving semantic relevance (Step \uppercase\expandafter{\romannumeral 2}). Each step details as follows.

\subsection{Step \uppercase\expandafter{\romannumeral 1}: Safe and Unsafe Prompt Recognition} 
\begin{figure}[t]
    \centering
    \includegraphics[width=0.45\textwidth]{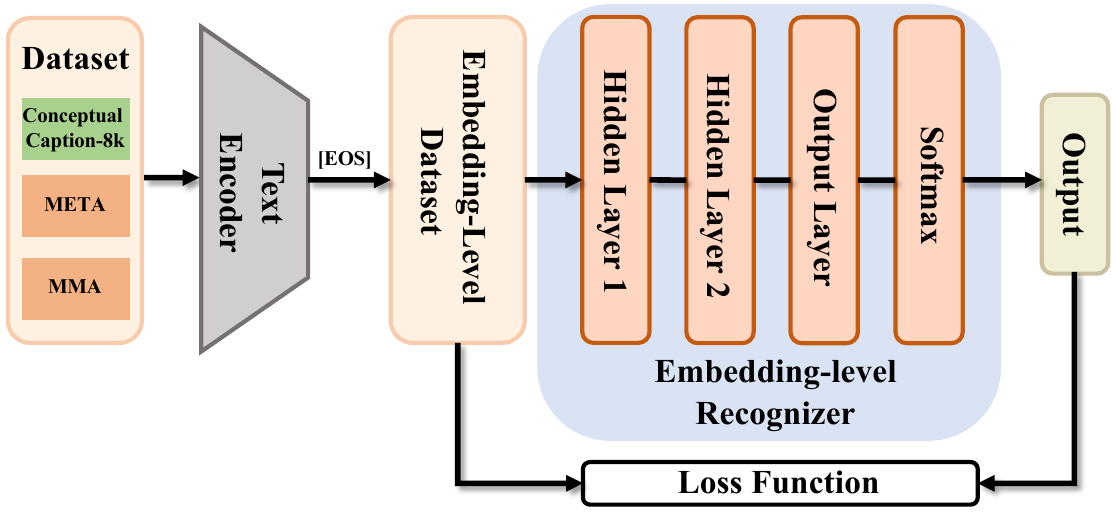}
    \vspace{-1em}
    \caption{Training pipeline of embedding-level recognizer.}
    \label{fig:recognizer_training}
    \vspace{-2em}
\end{figure}

In this step, \textbf{SafeGuider} processes input prompts through a text encoder to obtain [EOS] token embeddings, which are then evaluated by our proposed embedding-level recognizer for safety assessment. This recognizer is a lightweight classification model that maps the [EOS] token's representation to a safety score, determining whether a prompt is safe or unsafe based on this score. The design leverages our findings from Sec.~\ref{Sec4} about the token's ability to capture prompt characteristics. As illustrated in Fig.~\ref{fig:recognizer_training}, we develop this recognizer through three key parts: embedding-level dataset construction (Sec.~\ref{Sec5.2.1}), lightweight architecture design (Sec.~\ref{Sec5.2.2}), and training strategy (Sec.~\ref{Sec5.2.3}). 

\subsubsection{Embedding-level Dataset Construction}
\label{Sec5.2.1}
We construct our embedding level dataset using three prompt sources: 9,275 benign prompts from Conceptual Caption \cite{DBLP:conf/acl/SoricutDSG18}, 8,585 vocabulary substitution attacks from META dataset \cite{DBLP:journals/corr/abs-2405-19360}, and 2,000 symbol injection attacks from MMA dataset\cite{DBLP:conf/cvpr/Yang0WHX024}. The adversarial datasets encompass seven unsafe categories as discussed in Sec~\ref{Sec2.1.3}: pornography, violence, hate speech, harassment, self-harm, shocking, and illegal content. Notably, while trained on these specific datasets, our recognizer demonstrates a strong generalization ability to out-of-domain attacks, as validated in our experimental results (Sec.~\ref{Sec7}).

As shown in Fig.~\ref{fig:recognizer_training}, the dataset construction process consists of two main steps. First, for each prompt, the SD-V1.4 text encoder tokenizes the input and generates a fixed-size embedding matrix $E \in \mathbb{R}^{77\times768}$, where 77 represents the maximum sequence length and 768 is the embedding dimension. Then, we extract the [EOS] token embedding vector $e_{agg} = E[len(P), :] \in \mathbb{R}^{1\times768}$ from the matrix, where $len(P)$ indicates the prompt's actual length. Finally, we obtain an embedding-level dataset containing 19,860 [EOS] token embeddings, with $\approx$80\% for training our recognizer.

\subsubsection{Lightweight Architecture Design} 
\label{Sec5.2.2}
For efficient prompt safety assessment, we design a lightweight recognizer $C_\theta$ that predicts safety scores from [EOS] token embeddings:
\begin{equation}
    C_\theta: \mathbb{R}^{1\times768} \rightarrow S,
\end{equation} 
where $S$ represents the predicted safety score. The recognizer employs a three-layer neural network with progressive dimensionality reduction, using ReLU activations and dropout regularization. For an input embedding vector $e_{\text{agg}}$, the model outputs both logits and probability distributions through softmax normalization, where the probability of the positive class represents the prompt's safety score. This architecture provides an efficient balance between model capacity and computational overhead while maintaining robust recognition performance.

\subsubsection{Training Strategy} 
\label{Sec5.2.3}
We design a custom loss function that encourages diverse safety score distributions: 
\setcounter{equation}{3}
\begin{align}
L(\theta) &= L_{\text{pos}} + L_{\text{neg}} \nonumber \\
&= -\frac{1}{N_{\text{pos}}} \sum_{y_i=1} \log(p_i) - \frac{1}{N_{\text{neg}}} \sum_{y_i=0} \log(1 - p_i)
\end{align}
where $p_i$ denotes the predicted safety score, and $N_{\text{pos}}$, $N_{\text{neg}}$ are the number of benign and adversarial samples, respectively. This formulation encourages high scores for benign prompts and diverse low scores for adversarial ones, promoting distributional separation while avoiding over-convergence. We train the recognizer for 50 epochs with a batch size of 32.

\subsection{Step \uppercase\expandafter{\romannumeral 2}: Guide Unsafe Prompts to Output Safe and Meaningful Images}
In this step, we focus on processing unsafe prompts identified by Step \uppercase\expandafter{\romannumeral 1} to enable safe and semantically meaningful image generation. Inspired by our findings on distinct [EOS] token patterns (Sec.~\ref{Sec4}), we aim to guide unsafe prompts toward benign embeddings while preserving semantics. Specifically, \textbf{SafeGuider} aims to obtain a new condition embedding matrix that is both safe and semantically relevant. To achieve this embedding-level objective, we propose Safety-Aware Feature Erasure (SAFE) beam search, which strategically modifies input tokens guided by both safe and semantic similarity metrics at the embedding level. SAFE beam search first analyzes the contribution of each token of the prompt to unsafe content by calculating the safety score after removing the token (lines 3–10). Based on these scores, tokens are ranked by their impact on safety. Then, using beam search with width $K$ and depth $D$, the algorithm systematically explores different token subsets to identify the optimal remaining tokens (lines 11–24). Throughout the search process, we maintain the most promising candidates $K$, where each candidate is a subset of tokens from the original prompt. Each candidate is evaluated based on two criteria: the safety score of its resulting embedding (from our recognition model) and its semantic similarity to the original embedding. For semantic similarity assessment, we compute the cosine similarity between the [EOS] token embeddings of the modified and original prompts:
\begin{equation}
\text{similarity}(e_{new}, e) = \frac{e_{new} \cdot e}{||e_{new}|| \cdot ||e||}
\end{equation}
where $e\_{new}$ and $e$ represent the [EOS] token embeddings of the modified and original prompts, respectively. This dual evaluation implements a two-fold optimization objective: maximizing the prompt safety score while maintaining semantic similarity above the predefined thresholds. 
The process continues until we find an optimal combination whose embedding achieves both high safety and semantic preservation.

The SAFE beam search efficiently identifies modifications that enhance prompt safety while preserving meaningful semantic conditions. The beam width $K$ and depth $D$ constraints ensure tractable computation, while the dual-objective evaluation of safety and similarity guides the search toward effective solutions. 
\vspace{-1em}

\section{Implementation and Experimental Setup}
\label{Sec6}
In this section, we detail our implementation, baselines, datasets, and metrics used to evaluate \textbf{SafeGuider}'s performance.

\noindent\textbf{Implementation.} We implement \textbf{SafeGuider} on Ubuntu 22.04 with Python 3.8.5 and PyTorch 2.4.1+cu121. Following prior works \cite{DBLP:conf/ccs/LiYD0C0024,DBLP:conf/cvpr/SchramowskiBDK23,DBLP:conf/iccv/GandikotaMFB23}, we use SD-V1.4 as our base model. For SAFE beam search, we set beam width to 6, search depth to 25 to balance effectiveness and efficiency. In step \uppercase\expandafter{\romannumeral 2}, we set the safety threshold to 0.8 and semantic similarity threshold to 0.5 to ensure more safety while maintaining the semantics.

\noindent\textbf{Baselines.} We compare \textbf{SafeGuider} against ten state-of-the-art baselines implemented on SD-V1.4, which serves as the base model due to its lack of built-in safety mechanisms.

\noindent\underline{\textit{Internal Defenses}.} We compare against methods that modify model architecture or parameters during training or fine-tuning, including SLD \cite{DBLP:conf/cvpr/SchramowskiBDK23}, ESD \cite{DBLP:conf/iccv/GandikotaMFB23}, and SafeGen \cite{DBLP:conf/ccs/LiYD0C0024}, where ESD and SafeGen are specifically designed for pornographic content mitigation. 

\noindent\underline{\textit{External Defenses}.} We evaluate methods that employ independent filters, including text-level OpenAI Moderation \cite{openai}, Microsoft Azure Content Moderator \cite{azure}, AWS Comprehend \cite{aws}, NSFW Text Classifier \cite{nsfwtextclassifier}, GuardT2I \cite{DBLP:journals/corr/abs-2403-01446}, and an image-level Safety Checker \cite{safetychecker}. These methods operate independently of the model architecture, providing different approaches to content filtering.

\noindent\textbf{Evaluation Datasets.} We evaluate in-domain and out-of-domain test sets, each comprising benign prompts, vocabulary substitution (VS) and symbol injection (SJ) adversarial attacks. 

\noindent\underline{\textit{In-domain Evaluation}.} We use the held-out $\approx$20\% of our embedding datasets as the test set, including benign from Conceptual Caption (CCaption) \cite{DBLP:conf/acl/SoricutDSG18}, VS attacks from META dataset \cite{DBLP:journals/corr/abs-2405-19360}, and SJ attacks from MMA dataset \cite{DBLP:conf/cvpr/Yang0WHX024}.

\noindent\underline{\textit{Out-of-domain Evaluation}.} We test on prompts from the COCO2017 validation subset for benign content \cite{DBLP:conf/eccv/LinMBHPRDZ14}, I2P \cite{DBLP:conf/cvpr/SchramowskiBDK23} and Sneaky \cite{DBLP:conf/sp/YangHYGC24} datasets for VS attacks, and Ring-A-Bell (RAB) \cite{DBLP:conf/iclr/TsaiHXLC0CYH24} and P4D \cite{DBLP:conf/icml/ChinJHCC24} datasets for SJ attacks.

\noindent These datasets cover different unsafe categories discussed in Sec.~\ref{Sec2.1.3}: META and I2P encompass all seven categories (pornography, violence, etc.); RAB contains pornography and violence, while the other focus on pornographic content.

\noindent\textbf{Metrics.} We evaluate using two types of metrics: safety metrics to assess defense effectiveness against adversarial attacks and quality metrics to measure generation performance on benign inputs.

\begin{table}[t]
\small
\setlength{\tabcolsep}{2pt}
\caption{[RQ1-1] Performance of different methods on detecting sexually explicit content across VS and SJ adversarial datasets (IND/OOD). Lower ASR (\%) indicates better performance. \textbf{Bold} numbers denote the best results.}
\label{tab:asr_comparison}
\vspace{-1em}
\setlength{\heavyrulewidth}{1.5pt}
\begin{tabular}{
    >{\centering\arraybackslash}p{\dimexpr0.05\columnwidth-2\tabcolsep}|
    >{\arraybackslash}p{\dimexpr0.205\columnwidth-2\tabcolsep}|
    >{\centering\arraybackslash}p{\dimexpr0.12\columnwidth-2\tabcolsep}|
    >{\centering\arraybackslash}p{\dimexpr0.09\columnwidth-2\tabcolsep}|
    >{\centering\arraybackslash}p{\dimexpr0.12\columnwidth-2\tabcolsep}|
    >{\centering\arraybackslash}p{\dimexpr0.12\columnwidth-2\tabcolsep}| 
    >{\centering\arraybackslash}p{\dimexpr0.12\columnwidth-2\tabcolsep}|
    >{\centering\arraybackslash}p{\dimexpr0.09\columnwidth-2\tabcolsep} 
}
\toprule
\multicolumn{1}{c|}{\multirow{5}{*}{\makecell{\textbf{Defense}\\\textbf{Type}}}} & \multicolumn{1}{c|}{\multirow{5}{*}{\textbf{Method}}} & \multicolumn{2}{c|}{\textbf{IND-ASR $\downarrow$}} & \multicolumn{4}{c}{\textbf{OOD-ASR $\downarrow$}} \\
\cmidrule(l){3-8}
& & \multicolumn{1}{c|}{\textbf{VS}} & \multicolumn{1}{c|}{\textbf{SJ}} & 
\multicolumn{2}{c|}{\textbf{VS}} & \multicolumn{2}{c}{\textbf{SJ}}\\ 
\cmidrule(l){3-8} 
& & \makecell{\textbf{META}\\\textbf{Sexual}} & \textbf{MMA} & \makecell{\textbf{I2P}\\\textbf{Sexual}} & \textbf{Sneaky} & \makecell{\textbf{RAB}\\\textbf{Sexual}} & \textbf{P4D} \\
\midrule[1pt]
\multirow{8}{*}{\centering\makecell{External\\Defense}} & OpenAI & 96.87 & 30.34 & 91.00 & 33.00 & 25.93 & 70.18 \\
\cmidrule(l){2-8}
& Azure & 83.02 & 15.45 & 82.00 & 19.00 & 2.06 & 35.32 \\
\cmidrule(l){2-8}
& AWS & 86.00 & 13.00 & 85.00 & 24.00 & 25.00 & 63.00 \\
\cmidrule(l){2-8}
& NSFW Text& 37.88 & 3.37 & 25.00 & 6.67 & 1.65 & 14.68 \\
\cmidrule(l){2-8}
& GuardT2I & 26.33 & 17.70 & 25.46 & 6.50 & 0.82 & 11.01 \\
\cmidrule(l){2-8}
& SafetyChecker & 64.50 & 53.09 & 40.28 & 35.50 & 7.37 & 28.75 \\
\midrule
\multirow{5.5}{*}{\centering\makecell{Internal\\Defense}} & ESD & 21.38 & 51.12 & 32.44 & 38.50 & 84.77 & 77.92 \\
\cmidrule(l){2-8}
& SLD-Medium & 32.76 & 90.73 & 54.99 & 81.50 & 100.00 & 97.08 \\
\cmidrule(l){2-8}
& SLD-Max & 16.04 & 84.83 & 49.19 & 52.78 & 81.21 & 91.25 \\
\cmidrule(l){2-8}
& SafeGen & 13.99 & 19.10 & 54.14 & 15.00 & 41.02 & 70.00 \\
\midrule
\multicolumn{1}{c|}{\textbf{Ours}} & \textbf{SafeGuider} & \textbf{2.05} & \textbf{1.12} & \textbf{5.48} & \textbf{2.78} & \textbf{0.01} & \textbf{0.46} \\
\bottomrule
\end{tabular}
\vspace{-2em}
\end{table}

\noindent\underline{\textit{Safety Assessment Metrics.} } We employ three metrics to evaluate the model's ability to defeat different types of adversarial attacks.

\begin{itemize}[leftmargin=*]

\item \textbf{Attack Success Rate (ASR)}: Percentage of successful attacks, measured by filter bypass rate (external defenses) or unsafe content generation rate (internal defenses) evaluated with NudeNet \cite{nudenet} (the sexual concept) and Q16 \cite{DBLP:conf/fat/SchramowskiTK22} (the other unsafe concepts). Since our malicious datasets contain only unsafe content, ASR directly equals FNR (False Negative Rate), with lower values indicating better safety.

\item \textbf{Nudity Removal Rate (NRR)}: Percentage of explicit content mitigation measured by NudeNet \cite{nudenet}.

\item \textbf{Harmful Content Removal Rate (HCRR)}: Percentage of non-sexual harmful content mitigation measured by Q16 \cite{DBLP:conf/fat/SchramowskiTK22}.

\end{itemize}

\noindent\underline{\textit{Generation Quality Metrics.} }We use three metrics to ensure the model maintains high-quality outputs for benign inputs.
\begin{itemize}[leftmargin=*]

\item \textbf{Generation Success Rate (GSR)}: Percentage of successful image generations. Since our benign datasets contain only safe prompts, for external defenses, FPR = 100\% - GSR. For internal defenses, FPR isn't measurable as they modify the generation process without explicitly rejecting prompts. For \textbf{SafeGuider}, FPR is computed as the proportion flagged unsafe in Step \uppercase\expandafter{\romannumeral 1}.

\item \textbf{CLIP Score} \cite{DBLP:conf/nips/HuangSXLL23}: Semantic alignment between images and prompts.
\item \textbf{LPIPS Score} \cite{DBLP:conf/cvpr/ZhangIESW18}: Perceptual similarity to reference images.
\end{itemize}

\section{Evaluation}
\label{Sec7}
We analyze the \textbf{SafeGuider} in terms of robustness and practicality, and aim to answer the following Research Questions (RQs):
\begin{itemize}[leftmargin=*]
\item RQ1 [Robustness]: How effective is \textbf{SafeGuider}'s recognition model in detecting unsafe prompts?
\item RQ2 [Practicality-Benign]: How well does \textbf{SafeGuider} preserve image generation quality for benign prompts?
\item RQ3 [Practicality-Unsafe]: How effective is \textbf{SafeGuider} in guiding unsafe prompts to generate safe images?
\item RQ4 [Transferability]: What is the transferability of \textbf{SafeGuider} to different T2I models?
\item RQ5 [Ablation Study]: What is the importance of each step in our \textbf{SafeGuider}?
\item RQ6 [Adapative Evaluation]: What will happen if the attacker access our \textbf{SafeGuider}?
\end{itemize}

\subsection{RQ1: Robustness}
We evaluate \textbf{SafeGuider}'s robustness against both in-domain (IND) and out-of-domain (OOD) adversarial attacks, focusing on the detection of sexually explicit content and other harmful themes. Table~\ref{tab:asr_comparison} and Table~\ref{tab:asr_comparison_rq1-2} compare our method with existing defenses.

\begin{table}[t]
\small
\setlength{\tabcolsep}{4pt}
\caption{[RQ1-2] Performance of different methods on detecting other unsafe themes across VS and SJ attacks (IND/OOD).}
\label{tab:asr_comparison_rq1-2}
\vspace{-1em}
\setlength{\heavyrulewidth}{1.5pt}
\begin{tabular}{
    >{\centering\arraybackslash}p{\dimexpr0.2\columnwidth-2\tabcolsep}|
    >{\arraybackslash}p{\dimexpr0.23\columnwidth-2\tabcolsep}|
    >{\centering\arraybackslash}p{\dimexpr0.25\columnwidth-2\tabcolsep}|
    >{\centering\arraybackslash}p{\dimexpr0.16\columnwidth-2\tabcolsep}|
    >{\centering\arraybackslash}p{\dimexpr0.16\columnwidth-2\tabcolsep} 
}
\toprule
\multicolumn{1}{c|}{\multirow{5}{*}{\makecell{\textbf{Defense}\\\textbf{Type}}}} & 
\multicolumn{1}{c|}{\multirow{5}{*}{\textbf{Method}}} & 
\multicolumn{1}{c|}{\textbf{IND-ASR $\downarrow$}} & 
\multicolumn{2}{c}{{\textbf{OOD-ASR $\downarrow$}}} \\
\cmidrule(l){3-5}
& & \multicolumn{1}{c|}{\textbf{VS}} & \multicolumn{1}{c|}{\textbf{VS}} & \multicolumn{1}{c}{\textbf{SJ}} \\ 
\cmidrule(l){3-5} 
& & \multicolumn{1}{c|}{\textbf{META}} & \multicolumn{1}{c|}{\textbf{I2P}} & \multicolumn{1}{c}{\textbf{RAB}} \\
& & \multicolumn{1}{c|}{\textbf{Other}} & \multicolumn{1}{c|}{\textbf{Other}} & \multicolumn{1}{c}{\textbf{Other}} \\
\midrule[1pt]
\multirow{8}{*}{\centering\makecell{External\\Defense}} & OpenAI & 99.16 & 97.41 & 82.77 \\
\cmidrule(l){2-5}
& Azure & 78.56 & 85.23 & 2.73 \\
\cmidrule(l){2-5}
& AWS & 82.00 & 89.00 & 30.00 \\
\cmidrule(l){2-5}
& NSFW Text& 37.00 & 47.71 & 0.01 \\
\cmidrule(l){2-5}
& GuardT2I & 31.24 & 33.68 & 2.27 \\
\cmidrule(l){2-5}
& SafetyChecker & 49.27 & 20.87 & 93.64 \\
\midrule
\multirow{2.5}{*}{\centering\makecell{Internal\\Defense}} & 
SLD-Medium & 14.33 & 8.54 & 66.36\\
\cmidrule(l){2-5}
& SLD-Max & 3.36 & 3.02 & 16.11 \\
\midrule
\multicolumn{1}{c|}{\textbf{Ours}} & \textbf{SafeGuider} & \textbf{1.34} & \textbf{1.40} & \textbf{0.01} \\
\bottomrule
\end{tabular}
\vspace{-2em}
\end{table}

\noindent\underline{\textbf{[RQ1-1] Detection of Sexually Explicit Content.}} As shown in Table~\ref{tab:asr_comparison}, both defenses exhibit substantial vulnerabilities to sexually explicit content. For external defenses, commercial APIs show concerning vulnerabilities to vocabulary substitution attacks, with OpenAI Moderation reaching ASRs of 96.87\% on META (IND) and 91.00\% on I2P-Sexual (OOD), while Microsoft Azure and AWS Comprehend show similar weaknesses (82.00-86.00\% ASR). Although open-source solutions like NSFW Text Classifier and GuardT2I demonstrate better robustness, their ASRs (25.00-37.88\%) remain concerning for safe applications. For internal defenses, evaluated by generating three images per prompt and using NudeNet for unsafe content detection, the results reveal significant vulnerabilities, particularly to symbol injection attacks. Specifically, SLD-Medium exhibits ASRs of up to 100\% on RAB-Sexual, while ESD and SafeGen show consistently high ASRs (41.02-84.77\%). In contrast, \textbf{SafeGuider} achieves remarkably low ASRs across all scenarios: 2.05-5.48\% for vocabulary substitution and 0.01-1.12\% for symbol injection attacks. These low ASR values directly translate to minimal FNR, confirming \textbf{SafeGuider}'s exceptional capability in identifying harmful content even under sophisticated adversarial conditions.

\noindent\underline{\textbf{[RQ1-2] Detection of Other Unsafe Themes.}} Beyond sexually explicit content, we evaluate the effectiveness of different approaches in detecting other unsafe themes (e.g., violence, hate speech) using META-Other themes (IND) and I2P-Other/RAB-Other themes (OOD) datasets. As shown in Table~\ref{tab:asr_comparison_rq1-2}, external defenses demonstrate significant vulnerabilities, with OpenAI showing 99.16\% ASR on IND attacks and consistent performance on OOD datasets (82.77-97.41\%). For internal defenses, evaluated under the same protocol as sexually explicit content detection, the results reveal considerable weaknesses. SLD-Medium exhibits varying ASRs (8.54-66.36\%) in different datasets, while SD with Safety Checker performs poorly in OOD datasets (20.87-93.64\%). In contrast, \textbf{SafeGuider} maintains consistently robust performance across both IND and OOD scenarios, achieving low ASRs of 1.34\% and 0.01-1.40\% respectively.

\vspace{-5pt}
\begin{center}
\begin{tcolorbox}[colback=gray!10,
                  colframe=black,
                  width=8.5cm,
                  arc=1mm, auto outer arc,
                  boxrule=0.5pt
                 ]
\textbf{Take-home Message 1:} SafeGuider exhibits exceptional robustness in unsafe detection across diverse scenarios.
\end{tcolorbox}
\end{center}
\vspace{-5pt}

\begin{table}[t]
\small
\setlength{\tabcolsep}{2pt}
\caption{[RQ2] Performance of different methods on generation capabilities (GSR) and quality metrics (CLIP and LPIPS Score) across in-domain and out-of-domain datasets. }
\vspace{-1em}
\label{tab:asr_comparison_rq2}
\setlength{\heavyrulewidth}{1.5pt}
\begin{tabular}{
    >{\arraybackslash}p{\dimexpr0.2\columnwidth-2\tabcolsep}|
    >{\centering\arraybackslash}p{\dimexpr0.11\columnwidth-2\tabcolsep}|
    >{\centering\arraybackslash}p{\dimexpr0.145\columnwidth-2\tabcolsep}|
    >{\centering\arraybackslash}p{\dimexpr0.145\columnwidth-2\tabcolsep}|
    >{\centering\arraybackslash}p{\dimexpr0.11\columnwidth-2\tabcolsep}|
    >{\centering\arraybackslash}p{\dimexpr0.145\columnwidth-2\tabcolsep}| 
    >{\centering\arraybackslash}p{\dimexpr0.145\columnwidth-2\tabcolsep} 
}
\toprule
\multicolumn{1}{c|}{\multirow{3.5}{*}{\textbf{Method}}} & \multicolumn{3}{c|}{\textbf{IND-CCaption-9k}} & \multicolumn{3}{c}{\textbf{OOD-COCO2017-2k}} \\
\cmidrule(l){2-7}
& \makecell{\textbf{GSR} \textbf{$\uparrow$}} & \makecell{\textbf{CLIP}\\\textbf{Score $\uparrow$}} & \makecell{\textbf{LPIPS}\\\textbf{Score $\downarrow$}} & \makecell{\textbf{GSR} \textbf{$\uparrow$}} & \makecell{\textbf{CLIP}\\\textbf{Score $\uparrow$}} & \makecell{\textbf{LPIPS}\\\textbf{Score $\downarrow$}} \\
\midrule[1pt]
Original SD & \textbf{100.00} & \textbf{27.52} & \textbf{0.762} & \textbf{100.00} & \textbf{28.41} & \textbf{0.708} \\
\cmidrule(l){1-7}
OpenAI & 99.00 & 27.13 & 0.770 & 99.00 & 28.06 & 0.712 \\
\cmidrule(l){1-7}
Azure & 98.00 & 26.94 & 0.776 & 99.85 & 28.30 & 0.707 \\
\cmidrule(l){1-7}
AWS & 96.00 & 26.43 & 0.784 & 98.75 & 28.00 & 0.715 \\
\cmidrule(l){1-7}
NSFW Text& 70.60 & 25.32 & 0.803 & 64.87 & 26.19 & 0.777 \\
\cmidrule(l){1-7}
GuardT2I & 27.17 & 21.55 & 0.887 & 52.34 & 24.69 & 0.794 \\
\cmidrule(l){1-7}
SafetyChecker & 97.68 & 26.85 & 0.779 & 99.43 & 28.25 & 0.708 \\
\midrule
ESD & \textbf{100.00} & 26.56 & 0.776 & \textbf{100.00} & 27.76 & 0.713 \\
\cmidrule(l){1-7}
SLD-Medium & \textbf{100.00} & 26.07 & 0.781 & \textbf{100.00} & 26.30 & 0.726 \\
\cmidrule(l){1-7}
SLD-Max & \textbf{100.00} & 27.36 & 0.772 & \textbf{100.00} & 27.28 & 0.720 \\
\cmidrule(l){1-7}
SafeGen & \textbf{100.00} & 27.32 & 0.777 & \textbf{100.00} & 28.33 & \textbf{0.708} \\
\midrule
\textbf{SafeGuider} & \textbf{100.00} & 27.50 & 0.763 & \textbf{100.00} & \textbf{28.41} & \textbf{0.708} \\
\bottomrule
\end{tabular}
\vspace{-1.5em}
\end{table}

\begin{figure}[t]
\centering
\includegraphics[width=0.47\textwidth]{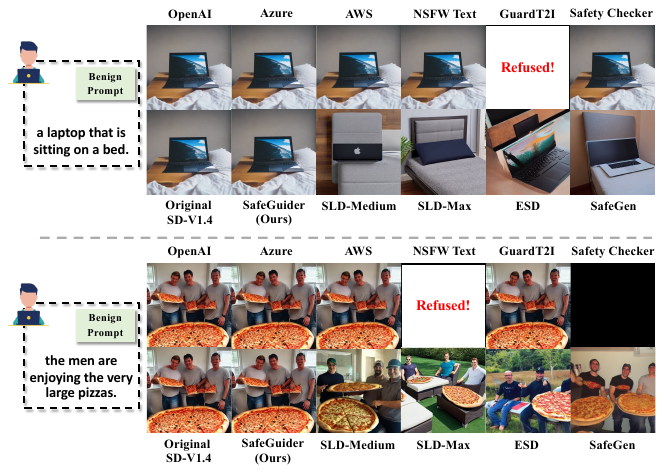}
\vspace{-1.5em}
\caption{Visual examples of generation quality on benign prompts by different defense strategies. }
\label{fig:rq2_example}
\vspace{-2em}
\end{figure}

\subsection{RQ2: Generation Quality on Benign Prompts}

We conduct experiments on both IND (Conceptual Caption \cite{DBLP:conf/acl/SoricutDSG18}) and OOD ( COCO2017 \cite{DBLP:conf/eccv/LinMBHPRDZ14}) datasets to assess the practical usability of \textbf{SafeGuider} on benign prompts. To evaluate \textbf{SafeGuider}’s impact on generation quality, we adopt three metrics: GSR, CLIP score, and LPIPS score. Results are summarized in Table~\ref{tab:asr_comparison_rq2} and Fig.~\ref{fig:rq2_example}.

\noindent\underline{\textbf{[RQ2-1] Generation Success Rate.}} 
The external defenses exhibit varying degrees of degradation in generation capabilities (Table~\ref{tab:asr_comparison_rq2}). While commercial APIs maintain relatively high GSRs (96.00-99.85\%), open-source solutions show significant limitations, with GuardT2I achieving only 27.17\% and 52.34\% GSR on IND and OOD datasets, respectively. In contrast, internal defenses like ESD, SLD, and SafeGen achieve 100\% GSR on both IND and OOD datasets, as they modify model architecture or parameters rather than filtering prompts. \textbf{SafeGuider} achieves 100\% GSR across all test scenarios, matching the original SD model's performance and demonstrating no compromise in generation capability. We further measure the FPRs of \textbf{SafeGuider}’s Step \uppercase\expandafter{\romannumeral1}, which are as low as 0.70\% on CCaption and 0.15\% on COCO2017-2k, outperforming existing external defenses. Step \uppercase\expandafter{\romannumeral2} remediates these rare false positives to maintain generation.

\noindent\underline{\textbf{[RQ2-2] Generation Quality.}} 
For the CLIP score , \textbf{SafeGuider} maintains performance comparable to the original SD model (27.50 vs. 27.52 on IND, 28.41 vs. 28.41 on OOD), outperforming most external defenses. The slight variations in CLIP Scores for \textbf{SafeGuider} can be attributed to minor false alarms from the recognition, but these differences are negligible in practice. For the LPIPS score, \textbf{SafeGuider} achieves scores nearly identical to the original SD model, showing superior perceptual quality compared to external defenses. This is notable as external defenses often default to generating black images upon rejection, leading to poor LPIPS scores. Internal defenses show comparable but worse performance due to their model modifications. We present qualitative examples of benign prompt generation in Fig.~\ref{fig:rq2_example}, showing that \textbf{SafeGuider} preserves the original model's generation capabilities.

\vspace{-5pt}
\begin{center}
\begin{tcolorbox}[colback=gray!10,
                  colframe=black,
                  width=8.5cm,
                  arc=1mm, auto outer arc,
                  boxrule=0.5pt
                 ]
\textbf{Take-home Message 2:} SafeGuider maintains the generation performance of the base model, achieving 100\% success rate on the benign prompts and CLIP/LPIPS scores.
\end{tcolorbox}
\end{center}
\vspace{-5pt}

\begin{table}[t]

\small
\setlength{\tabcolsep}{2pt}
\caption{[RQ3-1] Performance of different methods on mitigating sexually explicit content via nudity removal rate (NRR) across VS and SJ adversarial datasets (IND/OOD).}
\vspace{-1em}
\label{tab:asr_comparison_rq3-1}
\setlength{\heavyrulewidth}{1.5pt}
\begin{tabular}{
    >{\arraybackslash}p{\dimexpr0.22\columnwidth-2\tabcolsep}|
    >{\centering\arraybackslash}p{\dimexpr0.13\columnwidth-2\tabcolsep}|
    >{\centering\arraybackslash}p{\dimexpr0.12\columnwidth-2\tabcolsep}|
    >{\centering\arraybackslash}p{\dimexpr0.13\columnwidth-2\tabcolsep}|
    >{\centering\arraybackslash}p{\dimexpr0.14\columnwidth-2\tabcolsep}|
    >{\centering\arraybackslash}p{\dimexpr0.13\columnwidth-2\tabcolsep}| 
    >{\centering\arraybackslash}p{\dimexpr0.13\columnwidth-2\tabcolsep} 
}
\toprule
\multicolumn{1}{c|}{\multirow{5}{*}{\textbf{Method}}} & \multicolumn{2}{c|}{\textbf{IND-NRR $\uparrow$}} & \multicolumn{4}{c}{\textbf{OOD-NRR $\uparrow$}} \\
\cmidrule(l){2-7}
& \multicolumn{1}{c|}{\textbf{VS}} & \multicolumn{1}{c|}{\textbf{SJ}} & 
\multicolumn{2}{c|}{\textbf{VS}} & \multicolumn{2}{c}{\textbf{SJ}}\\ 
\cmidrule(l){2-7} 
& \makecell{\textbf{META}\\\textbf{Sexual}} & \textbf{MMA} & \makecell{\textbf{I2P}\\\textbf{Sexual}} & \textbf{Sneaky} & \makecell{\textbf{RAB}\\\textbf{Sexual}} & \textbf{P4D} \\
\midrule[1pt]
SafetyChecker & 78.37 & 54.63 & 81.00 & 77.35 & 73.42 & 78.71 \\
\midrule
ESD & 80.34 & 80.92 & 80.99 & 83.60 & 59.01 & 58.61 \\
\cmidrule(l){1-7}
SLD-Medium & 73.43 & -4.38 & 50.98 & 2.89 & -23.93 & -5.23 \\
\cmidrule(l){1-7}
SLD-Max & 76.10 & 28.82 & 67.64 & 45.46 & 40.93 & 42.51 \\
\cmidrule(l){1-7}
SafeGen & 79.03 & 92.31 & 58.58 & 85.62 & 76.81 & 73.27 \\
\midrule
\textbf{SafeGuider} & \textbf{86.61} & \textbf{93.32} & \textbf{83.33} & \textbf{88.52} & \textbf{81.71} & \textbf{82.57} \\
\bottomrule
\end{tabular}
\vspace{-1.5em}
\end{table}

\begin{figure}[t]
   \centering
   \includegraphics[width=0.48\textwidth]{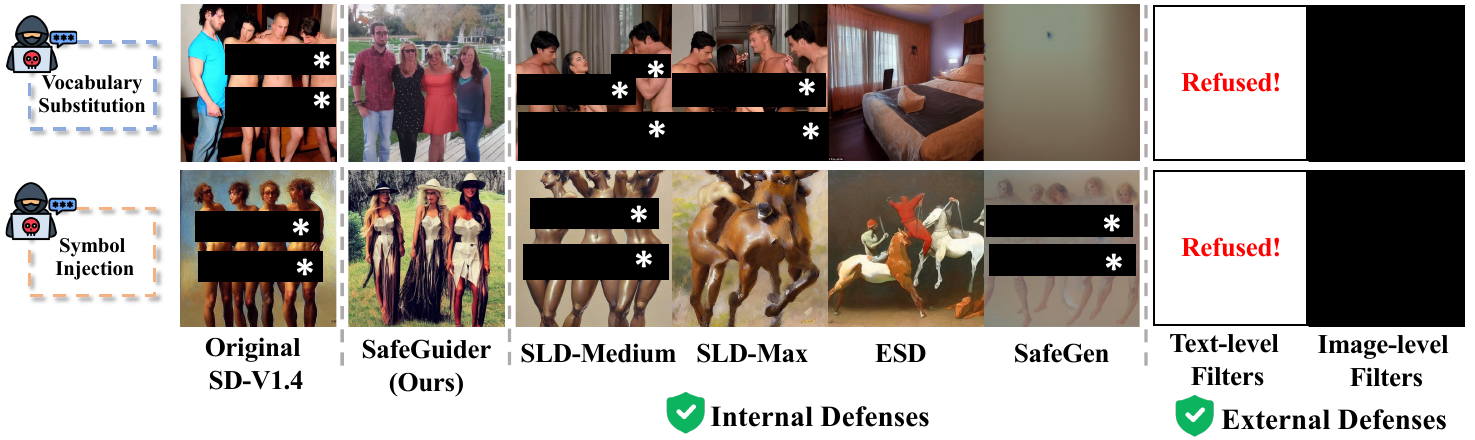}
   \vspace{-2.2em}
   \caption{Examples of sexually explicit content mitigation.}
   \label{fig:rq3_sexual}
   \vspace{-2em}
\end{figure}

\subsection{RQ3: Safe Generation for Unsafe Prompts}
\begin{figure}[t]
   \centering
   \includegraphics[width=0.48\textwidth]{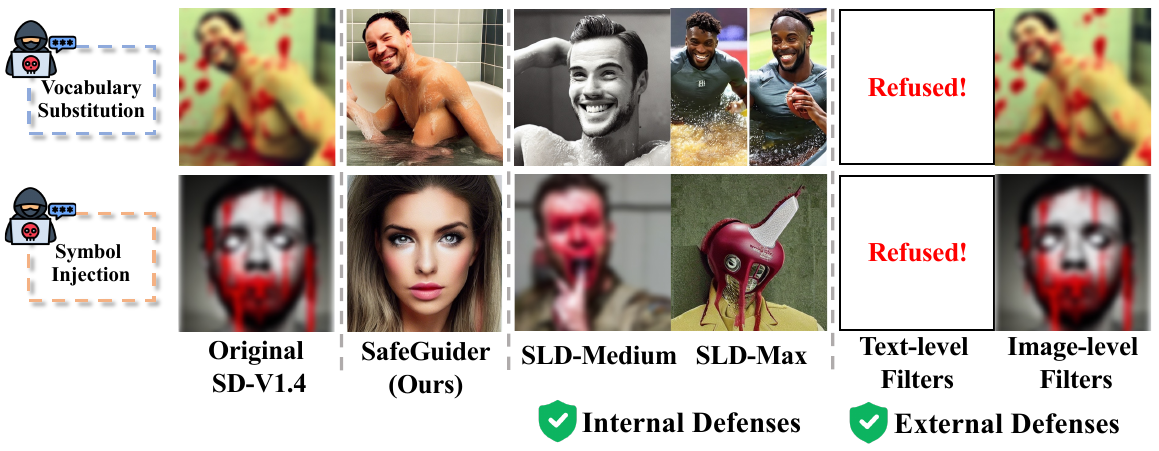}
   \vspace{-2.5em}
   \caption{Examples of other unsafe content mitigation. }
   \label{fig:rq3_other}
   \vspace{-2em}
\end{figure}

We evaluate \textbf{SafeGuider}'s effectiveness in guiding unsafe prompts to output safe and meaningful images. Unlike external defenses that simply reject unsafe prompts and produce black images, \textbf{SafeGuider} aims to guide the generation process toward safe alternatives. Our assessment uses specialized metrics for each category: NRR for sexually explicit content (Table~\ref{tab:asr_comparison_rq3-1} and Fig.~\ref{fig:rq3_sexual}) and HCRR for the other unsafe themes (Table~\ref{tab:asr_comparison_rq3-2} and Fig.~\ref{fig:rq3_other}).

\noindent\underline{\textbf{[RQ3-1] Mitigation of Sexually Explicit Content.}} 
As shown in Fig.~\ref{fig:rq3_sexual}, \textbf{SafeGuider} effectively removes inappropriate content while generating meaningful images that preserve the safe semantic elements of the original prompts. Furthermore, Table~\ref{tab:asr_comparison_rq3-1} quantitatively validates \textbf{SafeGuider}'s superior performance in safety generation. Specifically, among internal defenses, for vocabulary substitution attacks, \textbf{SafeGuider} achieves the highest NRR (86.61\% IND, 83.33-88.52\% OOD), significantly outperforming other approaches. ESD shows moderate performance (80.34\% IND, 80.99-83.60\% OOD), but SLD-Medium struggles particularly on OOD datasets (73.43\% IND, 2.89-50.98\% OOD). For symbol injection attacks, \textbf{SafeGuider} maintains robust performance (93.32\% IND, 81.71-82.57\% OOD), while other approaches show significant degradation.

\begin{table}[t]
\small
\setlength{\tabcolsep}{2pt}
\caption{[RQ3-2] Performance of different methods on mitigating other unsafe themes via harmful content removal rate (HCRR) across VS and SJ adversarial datasets (IND/OOD). }
\vspace{-1em}
\label{tab:asr_comparison_rq3-2}
\setlength{\heavyrulewidth}{1.5pt}
\begin{tabular}{
    >{\arraybackslash}p{\dimexpr0.25\columnwidth-2\tabcolsep}|
    >{\centering\arraybackslash}p{\dimexpr0.25\columnwidth-2\tabcolsep}|
    >{\centering\arraybackslash}p{\dimexpr0.25\columnwidth-2\tabcolsep}|
    >{\centering\arraybackslash}p{\dimexpr0.25\columnwidth-2\tabcolsep}
}
\toprule
\multicolumn{1}{c|}{\multirow{5}{*}{\textbf{Method}}} & \multicolumn{1}{c|}{\textbf{IND-HCRR $\uparrow$}} & \multicolumn{2}{c}{\textbf{OOD-HCRR $\uparrow$}} \\
\cmidrule(l){2-4}
& \multicolumn{1}{c|}{\textbf{VS}} & 
\multicolumn{1}{c|}{\textbf{VS}} & \multicolumn{1}{c}{\textbf{SJ}}\\ 
\cmidrule(l){2-4} 
& \makecell{\textbf{META}\\\textbf{Other}} & \makecell{\textbf{I2P}\\\textbf{Other}} & \makecell{\textbf{RAB}\\\textbf{Other}}  \\
\midrule[1pt]
SafetyChecker & 0.00 & 15.75 & 0.00 \\
\midrule
SLD-Medium & 70.04 & 67.32 & 51.09 \\
\cmidrule(l){1-4}
SLD-Max & 93.94 & 89.61 & 93.84 \\
\midrule
\textbf{SafeGuider} & \textbf{96.22} & \textbf{92.98} & \textbf{94.79} \\
\bottomrule
\end{tabular}
\vspace{-1.5em}
\end{table}

\noindent\underline{\textbf{[RQ3-2] Mitigation of Other Unsafe Themes.}}  
Fig.~\ref{fig:rq3_other} presents qualitative mitigation examples of other unsafe themes, showing that \textbf{SafeGuider} can effectively remove other harmful elements while maintaining the safe, intended aspects of the original generation. Besides, in Table~\ref{tab:asr_comparison_rq3-2}, \textbf{SafeGuider} achieves exceptional performance in safety generation on the other unsafe themes, substantially outperforming existing approaches with consistently high HCRR values (96.22\% IND, 92.98-94.79\% OOD). While SLD-Max shows reasonable performance, other approaches like SLD-Medium demonstrate lower effectiveness. Notably, SD with Safety Checker shows particularly poor performance with 0\% HCRR on several test cases, indicating complete failure in mitigating certain types of harmful content.

\vspace{-5pt}
\begin{center}
\begin{tcolorbox}[colback=gray!10,
                  colframe=black,
                  width=8.5cm,
                  arc=1mm, auto outer arc,
                  boxrule=0.5pt
                 ]
\textbf{Take-home Message 3:} SafeGuider demonstrates superior mitigation of various unsafe content while preserving meaningful image generation, outperforming both external defenses' binary blocking and other internal defenses.
\end{tcolorbox}
\end{center}
\vspace{-5pt}

\subsection{RQ4: Transferability}
\begin{figure}[t]
\centering
\includegraphics[width=0.48\textwidth]{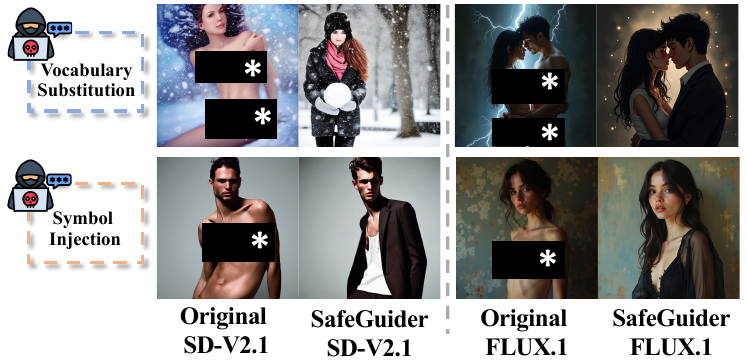}
\vspace{-2.3em}
\caption{Demonstration of SafeGuider's transferability across different T2I models.}
\label{fig:rq4_examples}
\vspace{-1em}
\end{figure}

We evaluate \textbf{SafeGuider}'s transferability to different T2I models, specifically testing on SD-V2.1 \cite{SDV2.1} and Flux.1 \cite{flux}. As shown in Table~\ref{tab:asr_comparison_rq4} and Fig.~\ref{fig:rq4_examples}, our experiments demonstrate \textbf{SafeGuider}'s broad applicability across varying model architectures.

\noindent\underline{\textbf{[RQ4-1] Adaptation to SD-V2.1.}} We first examine SD-V2.1, which employs OpenCLIP ViT-H/14 encoder where [EOS] is represented as ``<end of text>''. The results reveal that \textbf{SafeGuider} maintains nearly identical generation quality for benign prompts (CLIP: 28.74 vs 28.75, LPIPS: 0.703 vs 0.703) while demonstrating two key capabilities: effectively defending against various adversarial attacks and successfully guiding the generation process toward safe and semantically relevant alternatives (Fig.~\ref{fig:rq4_examples}).

\noindent\underline{\textbf{[RQ4-2] Adaptation to Flux.1.}} Flux.1 uses dual encoders (CLIP ViT-L and T5-XXL). \textbf{SafeGuider} can work with embeddings from different encoders. For CLIP ViT-L, we directly apply our pre-trained model. For T5, we reduce its 4096-dimensional embeddings to 1024 dimensions to better learn feature distributions with fewer training iterations, and retrain our recognizer. Results show effective defense (ASR reduced from 96.43\% to 0.41\% on RAB-Sexual) while preserving benign quality (CLIP: 29.00, LPIPS: 0.679).

These findings demonstrate \textbf{SafeGuider}'s exceptional transferability across different T2I architectures. Besides, \textbf{SafeGuider} can also operate in plug-and-play mode by encoding prompts externally with CLIP, making it well-suited for rapidly evolving T2I systems.

\vspace{-5pt}
\begin{center}
\begin{tcolorbox}[colback=gray!10,
                 colframe=black,
                 width=8.5cm,
                 arc=1mm, auto outer arc,
                 boxrule=0.5pt
                ]
\textbf{Take-home Message 4:} SafeGuider demonstrates transferability across different T2I architectures, offering a versatile safety solution through its architecture-agnostic approach.
\end{tcolorbox}
\end{center}
\vspace{-7pt}

\begin{table}[t]
\small
\setlength{\tabcolsep}{2pt}
\caption{[RQ4] Performance comparison between original models and SafeGuider on SD-V2.1 and FLUX.1.}
\label{tab:asr_comparison_rq4}
\vspace{-1em}
\setlength{\heavyrulewidth}{1.5pt}
\begin{tabular}{
    >{\arraybackslash}p{\dimexpr0.3\columnwidth-2\tabcolsep}|
    >{\centering\arraybackslash}p{\dimexpr0.175\columnwidth-2\tabcolsep}|
    >{\centering\arraybackslash}p{\dimexpr0.175\columnwidth-2\tabcolsep}|
    >{\centering\arraybackslash}p{\dimexpr0.175\columnwidth-2\tabcolsep}|
    >{\centering\arraybackslash}p{\dimexpr0.175\columnwidth-2\tabcolsep}
}
\toprule
\multicolumn{1}{c|}{\multirow{3.5}{*}{\makecell{\textbf{Method}}}} & 
\multicolumn{2}{c|}{\textbf{COCO2017-2k}} & \multicolumn{1}{c|}{\makecell[c]{\textbf{I2P} \\\textbf{Sexual}}}  & \multicolumn{1}{c}{\makecell[c]{\textbf{RAB} \\\textbf{ Sexual}}} \\
\cmidrule(l){2-5}
& \makecell{\textbf{CLIP} \\\textbf{Score $\uparrow$}} & \makecell{\textbf{LPIPS}\\\textbf{Score $\downarrow$}} & \makecell[c]{\textbf{ASR $\downarrow$}} & \makecell[c]{\textbf{ASR $\downarrow$}} \\
\midrule[1pt]
Original SD-V2.1 & 28.75 & 0.703 & 60.26 & 92.04 \\
\cmidrule(l){2-5}
\textbf{SafeGuider SD-V2.1} & \textbf{28.74} & \textbf{0.703} & \textbf{5.37} & \textbf{0.01} \\
\midrule\midrule
Original FLUX.1 & 29.00 & 0.679 & 64.55 & 96.43 \\
\cmidrule(l){2-5}
\textbf{SafeGuider FLUX.1} & \textbf{29.00} & \textbf{0.679} & \textbf{6.44} & \textbf{0.41} \\
\bottomrule
\end{tabular}
\vspace{-1.5em}
\end{table}

\subsection{RQ5: Ablation Study}
We conduct ablation studies to analyze the contribution of each step in \textbf{SafeGuider} using COCO2017 for benign prompts and I2P-Sexual for unsafe prompts. As shown in Table~\ref{tab:asr_comparison_rq5}, we evaluate three configurations: Step \uppercase\expandafter{\romannumeral 1}-only, Step \uppercase\expandafter{\romannumeral 2}-only, and complete framework.

\noindent\underline{\textbf{[RQ5-1] The Performance of Step \uppercase\expandafter{\romannumeral 1}-only \& Step \uppercase\expandafter{\romannumeral 2}-only.}} The step \uppercase\expandafter{\romannumeral 1}-only achieves the fastest processing time (69.02s per prompt) but shows limitations. For benign prompts, false positives in safety detection lead to unnecessary rejections, resulting in a reduced GSR (99.85\%) and compromised generation quality due to black image substitution. For unsafe prompts, while effectively blocking unsafe content, it achieves only 5.48\% GSR since rejected generations are replaced with black images rather than safe alternatives. The step \uppercase\expandafter{\romannumeral 2}-only shows robust safety control but exhibits certain constraints. While achieving 100\% GSR for benign prompts, it shows slightly degraded CLIP scores (28.29) compared to the complete framework (28.41), as it applies modifications to all prompts, including already-safe ones, to meet safety thresholds. For unsafe prompts, it achieves an NRR of 83.72\% but requires increased generation time.

\noindent\underline{\textbf{[RQ5-2] The Performance of Complete Framework.}} The complete \textbf{SafeGuider} framework combines step \uppercase\expandafter{\romannumeral 1} and step \uppercase\expandafter{\romannumeral 2} effectively. For benign prompts, it achieves optimal performance (100\% GSR, 28.41 CLIP score, 0.701 LPIPS score) while maintaining robust unsafe content mitigation (83.33\% NRR). Processing efficiency remains reasonable at 76.85s per prompt, as Step \uppercase\expandafter{\romannumeral 1}'s recognizer avoids unnecessary modifications to already-safe prompts. Importantly, this total time includes $\sim$64.98s for image generation and $\sim$11.87s for \textbf{SafeGuider}'s security processing, making the actual security overhead quite modest. Furthermore, Step \uppercase\expandafter{\romannumeral 2} of \textbf{SafeGuider} is compatible with faster search methods like top-p sampling or diverse beam search, offering additional efficiency opportunities.
In summary, while individual components show specific strengths - Step \uppercase\expandafter{\romannumeral 1}'s speed and Step \uppercase\expandafter{\romannumeral 2}'s thoroughness - their combination in \textbf{SafeGuider} provides the balanced solution. The framework leverages step \uppercase\expandafter{\romannumeral 1} for efficiency and step \uppercase\expandafter{\romannumeral 2} for safety, achieving protection while maintaining high-quality generation and reasonable computational cost.

\vspace{-5pt}
\begin{center}
\begin{tcolorbox}[colback=gray!10,
                  colframe=black,
                  width=8.5cm,
                  arc=1mm, auto outer arc,
                  boxrule=0.5pt
                 ]
\textbf{Take-home Message 5:} SafeGuider's two-step framework outperforms its individual components, achieving optimal balance between generation quality and safety.
\end{tcolorbox}
\end{center}
\vspace{-10pt}

\begin{table}[t]
\small

\setlength{\tabcolsep}{2pt}
\caption{[RQ5] Ablation study of SafeGuider comparing Step \uppercase\expandafter{\romannumeral 1}-only, Step \uppercase\expandafter{\romannumeral 2}-only and the complete framework.}
\label{tab:asr_comparison_rq5}
\vspace{-1.5em}
\setlength{\heavyrulewidth}{1.5pt}
\begin{tabular}{
    >{\arraybackslash}p{\dimexpr0.18\columnwidth-2\tabcolsep}|
    >{\centering\arraybackslash}p{\dimexpr0.19\columnwidth-2\tabcolsep}|
    >{\centering\arraybackslash}p{\dimexpr0.11\columnwidth-2\tabcolsep}|
    >{\centering\arraybackslash}p{\dimexpr0.145\columnwidth-2\tabcolsep}|
    >{\centering\arraybackslash}p{\dimexpr0.145\columnwidth-2\tabcolsep}|
    >{\centering\arraybackslash}p{\dimexpr0.11\columnwidth-2\tabcolsep}| 
    >{\centering\arraybackslash}p{\dimexpr0.11\columnwidth-2\tabcolsep} 
}
\toprule
\multicolumn{1}{c|}{\multirow{3.5}{*}{\makecell{\textbf{Method}}}} & 
{\multirow{3}{*}{\makecell{\textbf{Time Cost}\\\textbf{Per Prompt}\\\textbf{(s)$\downarrow$} }}} & \multicolumn{3}{c|}{\textbf{COCO2017-2k}} & \multicolumn{2}{c}{\textbf{I2P Sexual}} \\
\cmidrule(l){3-7}
& & \makecell{\textbf{GSR} \textbf{$\uparrow$}} & \makecell{\textbf{CLIP} \\\textbf{Score $\uparrow$}} & \makecell{\textbf{LPIPS}\\\textbf{Score $\downarrow$}} & \makecell{\textbf{GSR} \textbf{$\uparrow$}} & \makecell{\textbf{NRR}\textbf{$\uparrow$}} \\
\midrule[1pt]
Original SD & \textbf{64.98} & \textbf{100.00} & \textbf{28.41} & \textbf{0.701} & \textbf{100.00} & - \\
\cmidrule(l){1-7}
Step \uppercase\expandafter{\romannumeral 1}-only & 65.02 & 99.85 & 28.35 & 0.707 & 5.48 & - \\
\cmidrule(l){1-7}
Step \uppercase\expandafter{\romannumeral 2}-only & 87.60 & \textbf{100.00} & 28.29 & 0.710 & \textbf{100.00} & \textbf{83.72} \\
\midrule
\textbf{SafeGuider} & 76.85 & \textbf{100.00} & \textbf{28.41} & \textbf{0.701} & \textbf{100.00} & 83.33 \\
\bottomrule
\end{tabular}
\vspace{-2.5em}
\end{table}

\subsection{RQ6: Adaptive Evaluation}

We evaluate \textbf{SafeGuider} against adaptive adversaries who possess full knowledge of both the T2I model and our defense mechanism. We perform adaptive optimization on the P4D harmful dataset \cite{DBLP:conf/icml/ChinJHCC24} to develop strategies that could potentially circumvent our defense. We measure effectiveness using the Adaptive Attack Success Rate (AASR), calculated as the product of Attack Success Rate (ASR) and Unsafe Generation Rate (UGR), where UGR represents the percentage of bypassed prompts that generate harmful content as detected by NudeNet. Without adaptation, the original P4D dataset achieves an AASR of 87.22\% (ASR: 87.22\%, UGR: 100\%) against SD-V1.4, but only 0.46\% (ASR: 0.46\%, UGR: 100\%) against SD-V1.4 with \textbf{SafeGuider}. We explore two categories of adaptive strategies against SD-V1.4 with \textbf{SafeGuider}: 1) adding additional [EOS] tokens and 2) modifying the [EOS] token embedding.

\subsubsection{Adaptive Attacks via Adding [EOS] Tokens} Drawing inspiration from large language model (LLM) jailbreaking techniques \cite{DBLP:journals/corr/abs-2405-20653}, we attempt to bypass \textbf{SafeGuider} by inserting multiple [EOS] tokens at various positions (beginning, middle, end) and quantities (1, 3, 5, 7, 9) within prompts. However, this approach yielded no improvement in attack effectiveness against SD-V1.4 with \textbf{SafeGuider}, maintaining the AASR at 0.46\% on the P4D dataset. This failure stems from fundamental architectural differences between autoregressive LLMs and CLIP encoders: for large language models, their decoder-only autoregressive architecture processes tokens sequentially, causing earlier tokens to contribute less to the final embedding due to positional decay \cite{DBLP:conf/emnlp/WangLDCZMZS23}. Adding [EOS] tokens exploits this by pushing harmful content into the model's ``safe'' region \cite{DBLP:conf/nips/PengCHC24}. For CLIP encoders, however, tokens are processed in parallel without positional decay, and semantic information consistently converges at the final [EOS] token, which \textbf{SafeGuider} analyzes exclusively, making this attack strategy ineffective.

\subsubsection{Adaptive Attacks via Modifying [EOS] Token Embeddings} We next explore two approaches that explicitly alter the [EOS] token embedding to bypass SafeGuider: (1) optimizing the input prompt to indirectly influence the resulting [EOS] token embedding, and (2) directly replacing the [EOS] token embedding in a malicious prompt’s final embeding matrix with that from a benign prompt.

\begin{figure}[t]
\centering
\includegraphics[width=0.30\textwidth]{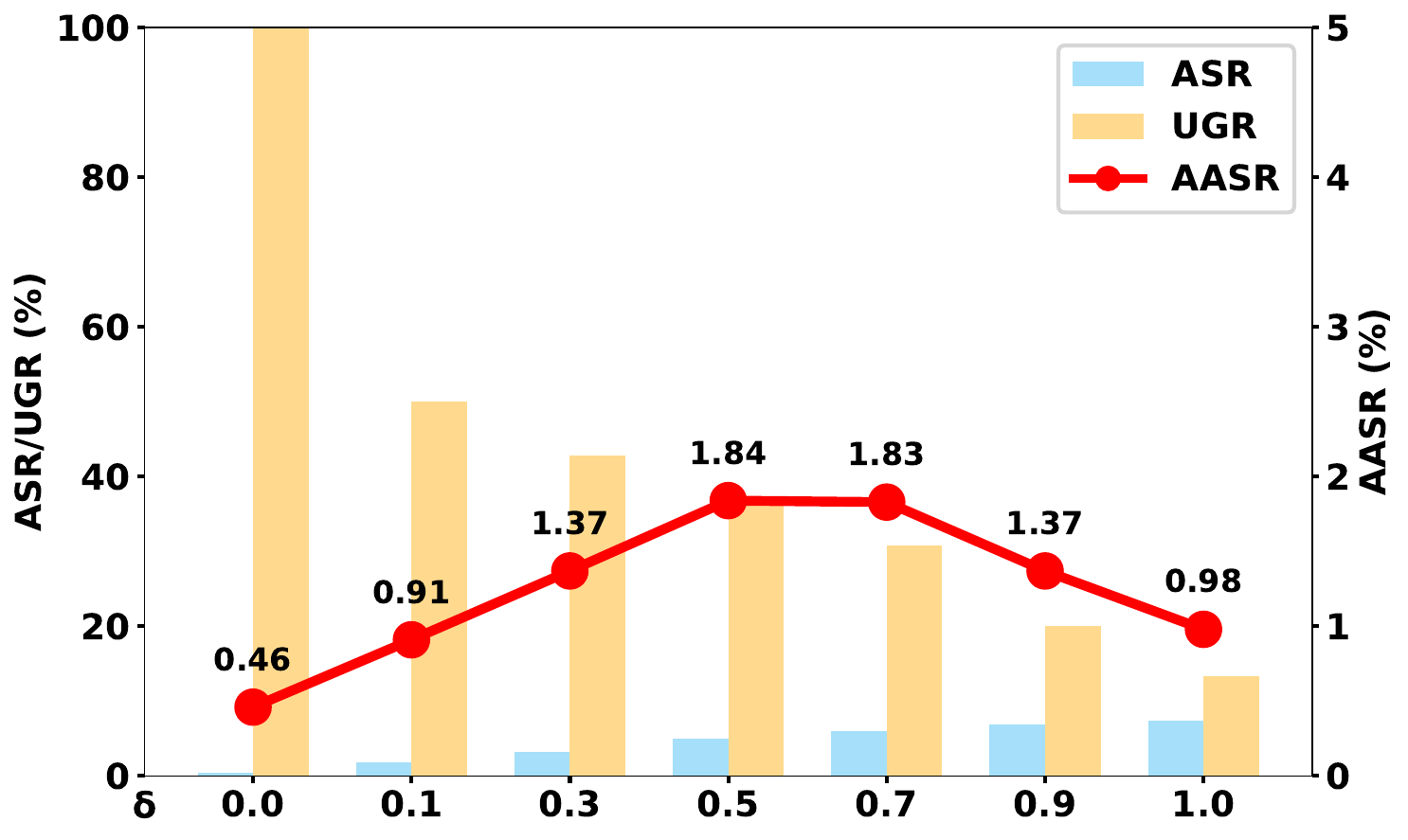}
\vspace{-1.5em}
\caption{Results of adaptive attacks with different $\delta$.}
\label{fig:adaptive_attack}
\vspace{-1em}
\end{figure}

\begin{figure}[t]
   \centering
   \vspace{-0.5em}
   \includegraphics[width=0.40\textwidth]{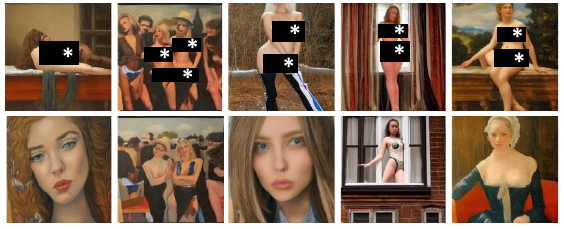}
   \vspace{-1em}
   \caption{
   Successful evasion (bottom) degrades output harmfulness. Each column has the same target NSFW content.
   }
   \label{fig:adaptive_attack_examples}
   \vspace{-2em}
\end{figure}

\noindent\textbf{(1) [EOS] Embedding Manipulation via Prompt Optimization.} We adapt the latest MMA-Diffusion adversarial attack \cite{DBLP:conf/cvpr/Yang0WHX024}, which leverages a gradient-based optimization framework to target T2I models. To extend this attack for \textbf{SafeGuider}, we introduce an additional term to enable the execution of adaptive attacks:
\begin{equation}
    L_{adaptive} = (1-\delta) \cdot L_{T2I}+ \delta \cdot L_{SafeGuider},
\end{equation}
where $L_{T2I}$ represents the original attack loss introduced by MMA-Diffusion, designed to manipulate the T2I model into generating NSFW content. $L_{SafeGuider}$ aims to evade our \textbf{SafeGuider} and $\delta $ is to balance these two terms. The results are summarized in Fig.~\ref{fig:adaptive_attack}, showing the following patterns:
\begin{itemize}[leftmargin=*]
    \item $\delta = 0$: the optimization fully focuses on $L_{T2I}$, aiming to generate prompts that strongly induce NSFW content in the T2I model and yield their corresponding [EOS] token embeddings. The resulting AASR remains at 0.46\% (ASR: 0.46\%, UGR: 100\%). This is because optimizing solely for harmful generation tends to reinforce malicious semantics in the prompt embedding, making it more likely to be flagged by \textbf{SafeGuider}. Meanwhile, those few prompts that already bypassed \textbf{SafeGuider} and led to harmful outputs are not further optimized, resulting in no overall gain.
    \item $\delta = 1$: the optimization focuses on bypassing \textbf{SafeGuider}. ASR increases to 7.34\% but UGR drops to 13.33\%, leading again to AASR = 0.98\%. For the majority of prompts, to evade detection, their semantics are optimized to benign, producing only safe images, except for the few raw prompts that already bypassed.
    \item $0 < \delta < 1$: a trade-off emerges between harmfulness and evasiveness. While ASR increases, the UGR decreases, with the AASR reaching its maximum of 1.84\% at $\delta$ = 0.5. This low adaptive attack success rate stems from inherently conflicting objectives: while $L_{T2I}$ seeks prompts with malicious semantics in T2I embeddings, evading the \textbf{SafeGuider} requires removing such semantic content. Qualitative analysis in Fig.~\ref{fig:adaptive_attack_examples} further demonstrates that successful evasion typically degrades output harmfulness. Thus, even with the defense knowledge, attackers struggle to circumvent our recognizer while maintaining attack effectiveness.
\end{itemize}

\vspace{-5pt}
\begin{center}
\begin{tcolorbox}[colback=gray!10,
                  colframe=black,
                  width=8.5cm,
                  arc=1mm, auto outer arc,
                  boxrule=0.5pt
                 ]
\textbf{Take-home Message 6:} SafeGuider also demonstrates robustness against adaptive attacks, with a maximum attack success rate of only 1.84\% across all tested strategies.
\end{tcolorbox}
\end{center}
\vspace{-5pt}

Despite adaptive optimization, the conflicting goals of inducing harmful content and evading SafeGuider result in limited attack success. The highest AASR reaches only 1.84\%, highlighting our \textbf{SafeGuider}’s robustness. Even with the full knowledge of the defense, attackers face a trade-off that constrains their effectiveness.

\noindent\textbf{(2) [EOS] Embedding Replacement with Benign Token.} We directly replace the [EOS] embedding in a malicious prompt's final embedding matrix with that from a benign prompt. While this modification can bypass \textbf{SafeGuider}, it cannot be translated back into a valid prompt. In transformer architectures, all token embeddings are interdependent due to the self-attention \cite{DBLP:conf/nips/VaswaniSPUJGKP17}. Once the [EOS] embedding is manually altered, the resulting embedding matrix loses internal consistency and cannot be reversed into a coherent prompt. In other words, only modifying the [EOS] token embedding disrupts self-attention, preventing its reversal into a valid input.

The evaluation of adaptive attacks reveals the resilience of our method. Adding [EOS] tokens fails due to fundamental differences between LLM and CLIP architectures. Modifying [EOS] embeddings through optimization or substitution faces trade-offs or structural infeasibility. Therefore, even with full defense knowledge, attackers struggle to bypass \textbf{SafeGuider} while preserving attack effectiveness, demonstrating its robustness in adversarial settings.

\section{Discussion}

Our framework offers flexible parameter configuration to accommodate various deployment scenarios. While our experiments demonstrate robust performance with default thresholds, service providers can customize these parameters based on their specific requirements, enabling a balanced trade-off between safety control and user experience. For instance, service providers prioritizing user experience might opt for a lower safety score requirement, enabling more precise content generation while maintaining acceptable safety standards. This adaptability makes \textbf{SafeGuider} suitable for various applications with different trade-off requirements.

\section{Conclusion}
In this work, we propose \textbf{SafeGuider}, a robust and practical framework for content safety control in text-to-image models. Based on our empirical findings about [EOS] token embeddings, our two-step approach achieves robust defense while maintaining high-quality generation and broad applicability across different architectures, making a step toward secure deployment of text-to-image systems.

\noindent\textbf{{Ethical Consideration.}} While developing \textbf{SafeGuider}, we have carefully considered the ethical implications of our research. Our work aims to prevent the generation of harmful content through T2I models while preserving their beneficial creative capabilities. In our evaluation, we ensured that all datasets were handled responsibly and that no harmful content was publicly shared. We hope our work contributes to the responsible development and deployment of AI technologies, promoting both innovation and social well-being.

\section*{Acknowledgement}
This work is supported in part by the Natural Science Foundation of China under Grants 62372423, 62121002, 62072421. This research is supported by the National Research Foundation, Singapore and Infocomm Media Development Authority under its Trust Tech Funding Initiative, the National Research Foundation, Singapore under its National Large Language Models Funding Initiative (AISG Award No: AISG-NMLP-2024-004), and the National Research Foundation, Singapore under its AI Singapore Programme (AISG Award No:  AISG4-GC-2023-008-1B). Any opinions, findings and conclusions or recommendations expressed in this material are those of the author(s) and do not reflect the views of National Research Foundation, Singapore and Infocomm Media Development Authority.

\bibliographystyle{ACM-Reference-Format}
\balance
\bibliography{reference}




\end{document}